\documentclass[11pt]{article}

\usepackage{setspace,graphicx,epstopdf,amsmath,amsfonts,amssymb,amsthm,versionPO} 
\usepackage{marginnote,datetime,enumitem,rotating,fancyvrb}
\usepackage{float}
\usepackage{natbib}
\usdate

\usepackage{booktabs}
\usepackage{pdflscape}
\usepackage{chngcntr}
\usepackage{longtable}
\usepackage{graphicx}
\usepackage{adjustbox}
\usepackage{multirow}
\usepackage{soul}            
\usepackage[format=plain,justification=justified,font=footnotesize]{caption,subfig}
\usepackage[capitalise,nameinlink,noabbrev]{cleveref}

\usepackage{titlesec}
\newcommand{\periodafter}[1]{#1.}
\titleformat{\subsubsection}[runin]
{\normalfont\bfseries}{\thesubsubsection}{1em}{\periodafter}

\excludeversion{notes}		
\includeversion{links}          

\iflinks{}{\hypersetup{draft=true}}

\ifnotes{%
\usepackage[margin=1in,paperwidth=10in,right=2.5in]{geometry}%
\usepackage[textwidth=1.4in,shadow,colorinlistoftodos]{todonotes}%
}{%
\usepackage[margin=1in]{geometry}%
\usepackage[disable]{todonotes}%
}

\makeatletter\let\chapter\@undefined\makeatother 

\usepackage{endnotes}    

\newtheorem{proposition}{Proposition}

\newtheorem{assumption}{Assumption}

\begin{document}

\setlist{noitemsep}  
\onehalfspacing      

\title{\bf Estimating the Impact of Social Distance Policy in Mitigating COVID-19 Spread with Factor-Based Imputation Approach\thanks{We are very grateful to Arthur H. O. van Soest (the editor), three anonymous referees, Victor Chernozhukov, Victor Couture, John Friedman, Jiti Gao, Yi He, Gael Martin, Yan Meng, Bin Peng, Donald Poskitt, Elie Tamer, Joakim Weill and the seminar participants in CEPR VoxEU, FEES virtual conference, International Symposium on Forecasting, International Association for Applied Econometrics Annual Conference, Asia Meeting of the Econometric Society, Australasia Meeting of the Econometric Society, Monash University, Placekey Community virtual conference for their helpful comments and discussions. Yanyi Ye gratefully acknowledges the financial support from the National Natural Science Foundation of China (Grant Number 72103017). Any mistakes are our own.}}

\author{
{\sc Difang Huang}\\
University of Hong Kong\vspace{0.2cm}\\
{\sc Ying Liang}\\
Toulouse School of Management\vspace{0.2cm}\\
{\sc Boyao Wu}\\
University of International Business and Economics\vspace{0.2cm}\\
{\sc Yanyi Ye}\\
Beijing University of Chemical Technology
}

\date{\vspace{-5ex}}

\maketitle
\thispagestyle{empty}

\clearpage
\pagenumbering{arabic}

\doublespacing

\begin{center}{\Large\bf Estimating the Impact of Social Distance Policy in Mitigating COVID-19 Spread with Factor-Based Imputation Approach}\end{center}

\vspace*{1in}

\centerline{\bf Abstract}
\medskip


We identify the effectiveness of social distancing policies in reducing the transmission of the COVID-19 spread. We build a model that measures the relative frequency and geographic distribution of the virus growth rate and provides hypothetical infection distribution in the states that enacted the social distancing policies, where we control time-varying, observed and unobserved, state-level heterogeneities. Using panel data on infection and deaths in all US states from February 20 to April 20, 2020, we find that stay-at-home orders and other types of social distancing policies significantly reduced the growth rate of infection and deaths. We show that the effects are time-varying and range from the weakest at the beginning of policy intervention to the strongest by the end of our sample period. We also found that social distancing policies were more effective in states with higher income, better education, more white people, more democratic voters, and higher CNN viewership.

\medskip



\clearpage

\section{Introduction}

	In response to the COVID-19 pandemic, governments worldwide have issued unprecedented policies restricting the movement of their citizens \citep[see][]{Chinazzi2020,Fang2020,Tian2020}. For example, in the US, the Federal Government issued social distancing recommendations and delegated health authority at the state level to implement the disease control procedures, leading to spatial and temporal variations in the implementation of social distancing measures \citep[see][among others]{Dave2020,Friedson2020,Gupta2020,Lauer2020,Simonov2020}. In this paper, we study the effectiveness of these social distancing policies and investigate the potential heterogeneous effects for states with distinct socio-demographic profiles.


	The traditional econometric approach may be invalidated by endogeneity concerns over the potential two-way feedback between social distancing and decline in virus infection, voluntary precautions, anticipation effects on health authority's mandates, and spillover effects between early and late states \citep{Chernozhukov2021,Goodman-Bacon_Marcus_2020,Manski2021}. To mitigate these concerns, we build a factor model with interactive fixed effects \citep{Bai2009,Bai2021} to clarify how social distancing policies dynamically influence the spread of COVID-19. We use the observed covariates to measure the individual's behavior that may change over time and use the interactive fixed effects to measure the unobserved common factors that drive the dynamics of infection distribution. These factors include the individual's awareness influenced by daily media reports and day-to-day changes in public expectations \citep[see][]{Fowler2020,Kucharski2020}. Our approach captures the direct effect of policies on the spread of COVID-19 and measures the indirect impact on human behavior. Besides, it also recognizes the individual's dynamic response to information and the voluntary change in her behavior \citep[see][]{Chernozhukov2021}.
	


	We derive the econometric specifications for both average and individual treatment effects to study social distancing policy effectiveness. Our specifications for the infection growth rate are structurally guided by the SEIR model while our econometric approach's validity may not rely on this model. We use the weekly growth rate of infection cases and deaths as the outcome of interest and focus on how these two variables are affected by the social distancing policy using state-level data. We consider the announcement date of stay-at-home orders to measure the social distancing policy and to avoid the complication in packaged mandate policies. We also measure the individual's behavior using the (lagged) population mobility data and use the number of testing and hospitalization as potential confounders. Our approach alleviates the endogeneity concern and provides the explicit identification of the relationship between social distancing policy and the reductions in infection growth rate.


Our key findings are as follows. We document that our model measures the cross-sectional and time-series variation of COVID-19 infections across all states of the United States. We further present evidence on the treatment effects of social distancing policy at the state level. During our sample period from February 20 to April 20, 2020, this non-pharmaceutical intervention across states effectively reduces the weekly growth rate in cases by 9.8\% and weekly growth rate in deaths by 7.0\%. The effects are time-varying, ranging from the weakest in the beginning few days to the strongest by the end of our sample period. The effects are cross-sectional disparate: states with higher income, higher education, more White people, more democratic voters, and higher CNN viewership experience a larger reduction in the infection growth rate than their counterparts. Our results are also robust to the choice of social distancing policy when considering the non-essential business closures as an alternative social distancing policy variable.



This paper extends our knowledge on several aspects. 
A growing strand of research in biostatistics, epidemiology, and economics is engaged in modeling the transmission of the COVID-19 pandemic \citep{Bao2021JFQA,Wu2023,ZHOU2024102528}. \cite{Li2021} provide an accurate forecast of the peak time of the epidemic based on the number of new cases and COVID-related deaths for 191 countries. \cite{Liu2021} apply Bayesian estimation models to forecast the daily infections for different countries. \cite{Korolev2021} builds the SEIRD epidemic model for the COVID-19 and identifies the basic reproduction number under the potential under-reporting of the number of cases. \cite{Kucharski2020} develop a mixed-effects non-linear regression model to estimate the evolution of daily growth of cases and COVID-related deaths using hospital data. \cite{lee2021sparse} propose the sparse HP filter to estimate the time-varying COVID-19 contact rate. \cite{ho2023go} extend \cite{Li2021} and \cite{Liu2021}, and apply Bayesian estimation with endogenously time-varying parameters for COVID-19 cases and death. Our model complements this strand of literature by combining the daily population movement and unobserved interactive fixed effects to explain the relative frequency and geographical distribution of new cases.

Another strand of literature focuses on the effectiveness of non-pharmaceutical interventions on COVID-19 infection \citep{CHEN2022126506,li2023impact}. Economists adopt the reduced-form approach (e.g., difference-in-difference, instrument variable, and synthetic control) and structural estimation approach to alleviate the endogeneity concern given health authorities endogenously implement such interventions. Governments across the world have implemented a wide range of non-pharmaceutical interventions (NPIs) to mitigate the spread of pandemic \citep{brauner2021inferring}. \cite{Chernozhukov2021} provide a causal structural estimation framework to study the dynamic effect of various US policies on the growth in infections and COVID-related deaths. It is also worth noting that \cite{bilgel2022effects} highlights a crucial intervention channel through which the social distancing policy influences social distancing behavior. The policy limits individuals' interactions during the pandemic outbreak, thereby effectively mitigating the widespread transmission of COVID infections. \cite{Cho2020}, \cite{Courtemanche2020}, and \cite{Fang2020} analyze the effectiveness of lockdown policies in Sweden, US, and China, respectively. \cite{Hsiang2020} use the panel data approach to study the effect of government interventions on the growth rate of infection cases in China, France, Iran, Italy, South Korea, and the US. \cite{Qiu2020} implement the instrumental variable approach to study the role of various interventions in mediating the transmission of COVID-19 in China. \cite{callaway2023policy} evaluate the effects of government policies in response to the COVID-19 pandemic and argue that unconfoundedness type approaches are more useful for policy evaluation than difference-in-differences methods.

Recent works in epidemiology employ Monte Carlo simulations to study the dynamics of cases and deaths under various policy recommendations. The uncertainty on the values of key model parameters and the strong assumptions on the effect of social distancing policy may lead to inaccurate predictions about outcomes and policy counterfactuals. Moreover, simulated models do not consider that people may voluntarily take precautionary behavior in response to increased risk. \cite{HORTACSU2021} study the implications of policy recommendations for the infection growth of COVID-19 using US data in the SIR (Susceptible-Infectious-Removed) framework. \cite{Fernandez-Villaverde2020} use data on COVID-related deaths in various cities to forecast the future infection growth rate under various policy interventions using the SIRD model. \cite{Pei2020} use county-level data of cases and deaths with population mobility data to conduct Monte Carlo simulations of government policies; they find that these policies can reduce the number of cases and deaths by more than 50\% 1--2 weeks following implementation. \cite{julliard2023spread} use a stochastic version of the workhorse SIR epidemiological model to account for spatial dynamics generated by network interactions and find that the UK lockdown measure reduced total propagation by 44\%. We contribute to this strand of literature by adopting the factor imputation approach, which accounts for the observed and unobserved factors that could affect the cases and deaths over time, and use cross-sectional data to consistently estimate the treatment effects of social distancing across all states in the US.

This paper also provides insights into understanding how human behavior may affect the effectiveness of disease prevention policies. Some studies show that such policies' effectiveness depends on state-level characteristics. \cite{Adolph2020} find that political partisanship explains the Governors' policy-making related to social distancing and shows that a delay in issuing the mandate is likely to significantly harm public health. \cite{Baud2020} document that the social distancing responses vary by county characteristics, including political partisanship, media consumption, and racial and ethnic composition. \cite{Dave2020} using state-wide data for Texas (US), show that counties with early shelter-in-place (referred to elsewhere as stay-at-home) orders are comparatively more effective than counties issuing orders late. \cite{Simonov2020} study the causal effect of Fox News consumption on individuals' stay-at-home behaviors and show that Fox News consumption was associated with non-compliance with social distancing policies during the crisis. \cite{Weill2020} show that an individual's income strongly differentiates their response to state-level emergency declarations. \cite{chen2023efficient} propose a COVID-19 epidemic control framework for efficient social distancing policies minimizing disease spread and economic risks. We show the effectiveness of social distancing in reducing the infection of COVID-19 and demonstrate the extent to which socio-demographic characteristics, including income, race, education, political beliefs, and media viewership, measured at a state level, are related to variations in the effectiveness of social distancing. We highlight the importance of taking this heterogeneity into account in future policy-making.
The remainder of the paper is structured as follows. \Cref{section:data} describes the data. \Cref{section:model} introduces our model. \Cref{section:results} presents the empirical results. \Cref{section:discuss} discusses the results. We show the detailed model in \Cref{section:appendix_detailed model}, proofs in \Cref{section:appendix_proof}, additional results in \Cref{section:appendix_results}, the SEIR model in \Cref{section:appendix_model}, and other details in \Cref{section:appendix_details}.

\section{Data}\label{section:data}

	This study employs two primary datasets to investigate the severity of the COVID-19 pandemic and the efficacy of social distancing policies. The first dataset, obtained from The New York Times (NYT), provides daily US state-level COVID-19 cases and deaths data. To ensure data completeness, we supplement missing values with data from the JHU CSSE and the Covid Tracking Project, which document the number of tests and hospitalizations. The sample period is restricted to 20 February and 20 April 2020 due to the inconsistent reporting of COVID-19 cases and deaths by many states prior to mid-February and the relaxation of social distancing policies by numerous states on 20 April 2020.\footnote{We choose the sample period for two reasons. First, many states are not reporting the cases and deaths of COVID-19 until the middle of February \citep{Manski2021}, we use the data from February 20 and also cross-validate the data from NYT with the data from JHU CSSE and from Covid Tracking Project to solve the miss reporting issue in the historical data. Second, many states reopened the economy starting from April 20, and therefore, we use these two dates as the beginning and end dates.}

	The second dataset, sourced from \cite{Adolph2020}, documents the implementation of state-level social distancing policies, including restrictions on gatherings, restaurants, school closures, non-essential business closures, and stay-at-home orders (detailed in \Cref{section:appendix_details}). As state governments often implemented multiple measures simultaneously to combat the pandemic, this study focuses on the most restrictive policy -- stay-at-home orders -- to avoid complications associated with analyzing multiple concurrent measures.\footnote{We focus on the announcement date for these policies instead of mandate date for two reasons. First, the state government's policies depend on the voluntary compliance of citizens and therefore the effects of these measures should be immediate given the threat of sanction. Second, our goal is to estimate the treatment effects of social distancing, which naturally points to announcement dates. Moreover, the dates for the announcement and implementation are often closely related.} The robustness of the results is confirmed using different timing and policy measures, as discussed in \Cref{section:appendix_results}. During the sample period, California was the first state to announce a stay-at-home order on 19 March 2020, with 39 out of 55 states following suit within the period.

	To further enhance the analysis, we acquire state-level population mobility data from Google Mobility, SafeGraph, and PlaceIQ, as well as state-level socio-demographic data, including income, race, education, political preference, and media consumption from the Bureau of Economic Analysis, Community Survey of the Census Bureau, PlaceIQ, MIT Election Lab, and Nielsen Local TV View.




%
%
%

\section{Econometric Methods}\label{section:model}

\subsection{Notation}

We consider the weekly growth rate of cases and deaths as the outcome variables $Y_{it}$:
\begin{equation}
Y_{i t} = 
\begin{cases}
	&\Delta \log \left(C_{i t}\right):=\log \left(\Delta C_{i t}\right)-\log \left(\Delta C_{i, t-7}\right) \\
	&\Delta \log \left(D_{i t}\right):=\log \left(\Delta D_{i t}\right)-\log \left(\Delta D_{i, t-7}\right) \\
\end{cases}       
\end{equation}
where we use either the weekly growth rate of cases or deaths as the outcome variables $Y_{i t}$ for  state $i$ at day $t$. $C_{i t}$ is the cumulative confirmed cases for state $i$ at day $t$, $D_{i t}$ is the cumulative confirmed COVID-related deaths for state $i$ at day $t$, $\Delta$ is the differencing operator. Following \cite{Chernozhukov2021}, we choose these variables as they are the key metrics for policymakers when deciding COVID-related policies as they smooth daily idiosyncratic fluctuations and periodic fluctuations arising on certain days of the week.

We denote the treated group with $\mathcal{T}$ who received the social distancing policy during the observation window and the control group that had not been exposed to the treatment with $\mathcal{C}$. The group size for treated group and control group are $N_{1}$ and $N_{0}$ where the total group size $N = N_{1} + N_{0}$. We denote the total observation period as $T$, and $T_{0,i}$ being the pre-treatment period for state $i$. As the main objective is to study whether social distancing policy affects the transmission of the disease, we denote $Y_{it}(\cdot)$ to distinguish whether state $i$ implemented the social distancing policy in time $t$ \citep{RUBIN1976}. Specifically, $Y_{it}(1)$ and $Y_{it}(0)$ denote the weekly growth rate in new cases and COVID-related deaths in state $i$ and time $t$ with and without a social distancing policy, respectively.

The policy intervention effect to state $i$ at time $t$ is captured by
\begin{equation}
\theta_{it} = Y_{it}(1) - Y_{it}(0).
\end{equation}

As we do not simultaneously observe $Y_{it}(1)$ and  $Y_{it}(0)$, the observed data can be shown as
\begin{equation}\label{equation:rubin}
Y_{it} = \mathcal{D}_{it}Y_{it}(1) + (1 - \mathcal{D}_{it}) Y_{it}(0),
\end{equation}
where $\mathcal{D}_{it}$ is the dummy variable that equals 1 if the state $i$ is under social distancing (treatment) at time $t$, and 0 otherwise. We have the model for observed $Y_{it}$:
\begin{align}
	Y_{it} = \theta_{it} \mathcal{D}_{it} + X_{i,t-m}^{'} \beta + Z_{i,t}^{'} \gamma + u_{it} \;\; \mathrm{with} \;\; u_{it} = \Lambda_{i}^{'} F_t + e_{it}, 
	\tag{4}
\label{equation:combine}
\end{align}
where $\theta_{i t}$ is the heterogeneous treatment effect that captures the impact of social distancing on weekly growth rate in new cases (deaths) for state $i$ at day $t$, $X_{i,t-m}$ is a vector of observed covariates with the lag of $m$ days to reflect the time lag between the time of infection and case (death) confirmation, include the historical population mobility measures that are epidemiologically informative to characterize the further distribution of infection cases both nation-wide and state-wide \citep[see][for examples]{Baud2020,Chinazzi2020,Kraemer2020}.\footnote{Follow the \cite{Chernozhukov2021}, for the growth rate in infection cases, we choose the lag $m = 14$ to capture the time delay between infection and confirmation and choose the lag $m = 21$ to capture the time delay between exposure and deaths.}  $Z_{i,t}$ is a vector of observed covariates that are not affected by the social distancing policy yet related to the distribution and growth in new cases, including the number of tests and the number of hospitalizations. Besides, $ u_{it} $ has an unknown factor structure, $F_{t}$ is a vector of common latent factors that drives the distribution of new infections across all states to change over time, $\Lambda_{i}$ is the factor loading vector, $C_{it} := \Lambda_{i}^{\prime} F_{t}$ is the interactive fixed effect, $e_{i t}$ is the error term, and $ \Lambda_{i} $, $ F_t $ and $ \varepsilon_{it} $ are all unobserved.\footnote{For the growth rate in new cases, we choose the lag $m = 14$ to capture the time delay between infection and confirmation and the lag $m = 21$ to capture the time delay between exposure and death. We include the historical population mobility measures that are epidemiologically informative to characterize the further distribution of new  cases both nationwide and state-wide. We present the proofs in the \Cref{section:appendix_proof}.} The common factors may capture differential time trends, as we illustrated in \Cref{section:appendix_assumption}.


We can represent the observed data of $Y$ as
\begin{equation*}
Y=\left[\begin{array}{cc}
	Y(0)_{T_{0} \times N_{0}} & Y(0)_{T_{0} \times N_{1}} \\
	Y(0)_{T_{1} \times N_{0}} & Y(1)_{T_{1} \times N_{1}} 
\end{array}\right].
\end{equation*}

As $Y_{it}(0)$ is not observable for state $i \in \mathcal{T}$ and $t \geq T_{0,i}+1$, we can represent the observed data of $Y(0)$ as
\begin{equation*}
Y(0)=\left[\begin{array}{cc}
	Y(0)_{T_{0} \times N_{0}} & Y(0)_{T_{0} \times N_{1}} \\
	Y(0)_{T_{1} \times N_{0}} & \textsc{Miss}
\end{array}\right],
\end{equation*}
where we denote the block matrix $Y(0)_{T_{0} \times N_{0}}$ with $\textsc{Bal}$, the block matrix $[ Y(0)_{T_{0} \times N_{0}}~Y(0)_{T_{1} \times N_{0}}]'$ with $\textsc{Tall}$, the block matrix $[Y(0)_{T_{0} \times N_{0}}~Y(0)_{T_{0} \times N_{1}}]$ with $\textsc{Wide}$, and  block matrix $\textsc{Miss}$. Therefore, we need to estimate the counterfactual distribution for $\textsc{Miss}$ block.

The model provides a valid estimation (inference) for average and heterogeneous treatment effects without imposing the parallel trend and error distribution assumptions \citep{Bai2021} and alleviates the endogeneity concerns when applying either a difference-in-differences or an instrumental variable approach (see more discussion on \Cref{section:appendix_large sample}). 

\subsection{The Model}
As detailed above, our model estimates average and individual treatment effects in a panel data setting with interactive fixed effects. Under a set of assumptions on the factors, factor loadings, error terms, and order conditions, as described in \Cref{section:appendix_assumption}, our model employs an algorithm that first estimates the coefficients using the control group observations. It then estimates the latent factors and factor loadings using the residuals in a $\textsc{Tall}$ block matrix and a two-step approach, respectively. We estimate the missing values using the estimated coefficients, factors, and factor loadings. Finally, we compute the individual and average treatment effects for the treated group and subgroups within the treated group. Readers may refer to \Cref{section:appendix_estimation} for more details on the estimation procedure.

Our model establishes the asymptotic normality of the average treatment effects on the treated group and the heterogeneous treatment effects on subgroups within the treated group. We can consistently estimate the asymptotic variances using our proposed estimators. The large sample properties of the estimators are derived under a set of assumptions that ensure the identification of the factor structure, weak correlation of error terms, and the validity of the central limit theorems for the factors and factor loadings. More details on these properties can be found in \Cref{section:appendix_large sample}.

We further discuss the validity of the assumptions used when applying our model to the state-level panel data. We create a social distancing policy index $\mathcal{D}_{i,t}$ for each state, which equals one if the stay-at-home order is announced in state $i$ on day $t$ and zero otherwise. Using stay-at-home orders as the measure for social distancing, we have $N_{0}$ = 16, $T_{0}$ = 28, $N$ = 55, and $T$ = 61. Therefore, Assumption 2, which states that $\frac{\sqrt{N}}{\min \left\{N_{0}, T_{0}\right\}} \rightarrow 0$ and $\frac{\sqrt{T}}{\min \left\{N_{0}, T_{0}\right\}} \rightarrow 0$ as $N_{0}, N \rightarrow \infty$ and $T_{0}, T \rightarrow \infty$, is satisfied for the state-level data we considered. As a robustness check, we also consider non-essential business closures as a social distancing policy, with $N_{0}$ = 24, $T_{0}$ = 28, $N$ = 55, and $T$ = 61, as discussed in \Cref{section:appendix_results}.

\subsection{Comparisons with Other Methods}

Our model provides a robust estimate and inference for treatment effects without relying on restrictive distributional assumptions for error terms or parallel trend assumptions for average outcomes of treated and control units in the absence of treatment \citep{Bai2021}. In the context of endogenously enacted social distancing policies, recent research has employed difference-in-differences or instrumental variable approaches to address endogeneity concerns \citep{brauner2021inferring,callaway2023policy,Cho2020}.

However, these studies may overlook crucial observed and unobserved factors that confound the relationship \citep{Chernozhukov2021,Courtemanche2020,Fang2020} . Reverse causality exists between social distancing and the growth in new cases, as governments introduce restrictions in response to increases in cases and deaths \citep{Cho2020}. Voluntary precaution effects, whereby people take precautions before government restrictions are announced, can lead to biased difference-in-differences estimates \citep{Qiu2020}. Anticipation effects of policy announcements and spillover effects from other states can further bias the estimates \citep{Hsiang2020}.

In contrast, our factor model approach consistently estimates the average and heterogeneous treatment effects by identifying the causal relationship between social distancing and reductions in COVID-19 infections while accounting for factors that influence virus transmission over time and across states \citep{bai2002determining,Bai2009}. Our method traces the policy's effect on each unit before and after implementation, addressing endogeneity concerns by capturing reverse causality and voluntary precaution effects. By controlling for observed confounders and interactive fixed effects that remove partial correlation varying across units and time \citep{bai2019rank,Bai2021}, our approach ensures unbiased estimates that accurately reflect the true causal impact of social distancing policies on the spread of COVID-19.

\section{Empirical Results}\label{section:results}

%
%

As demonstrated in our SEIR model in \Cref{section:appendix_model}, social distancing interventions impact infection distribution both directly and indirectly. These measures not only reduce physical interaction between infected and susceptible individuals, but also indirectly influence virus transmission by raising public awareness. Our model accounts for these factors, offering a valuable tool for estimating new infection incidence across states. This information aids policy-makers in making statistically accurate assessments of COVID-19's future growth and spread patterns.\footnote{We conduct our analysis using the R statistical software, employing the associated fbi package (https://github.com/cykbennie/fbi) specifically designed for factor-based imputation data analysis.}

Social distancing requirements affect COVID-19 transmission directly and indirectly. By reducing physical interaction between infected and susceptible individuals, direct effects are achieved. Indirectly, these measures raise awareness of the virus among the population. Our model incorporates these effects, providing a practical tool for estimating new infection incidence in states and assisting policy-makers in conducting statistically accurate assessments of future COVID-19 growth and dissemination patterns.

\subsection{Average Treatment Effects of Social Distancing}

\begin{figure}[htbp]
	\centering
	\subfloat{\includegraphics[scale=0.55]{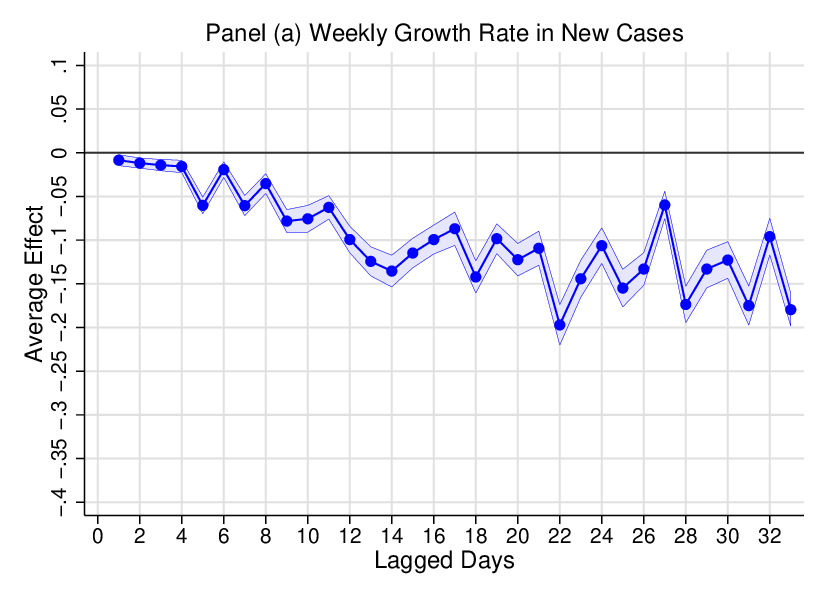}}
	\subfloat{\includegraphics[scale=0.55]{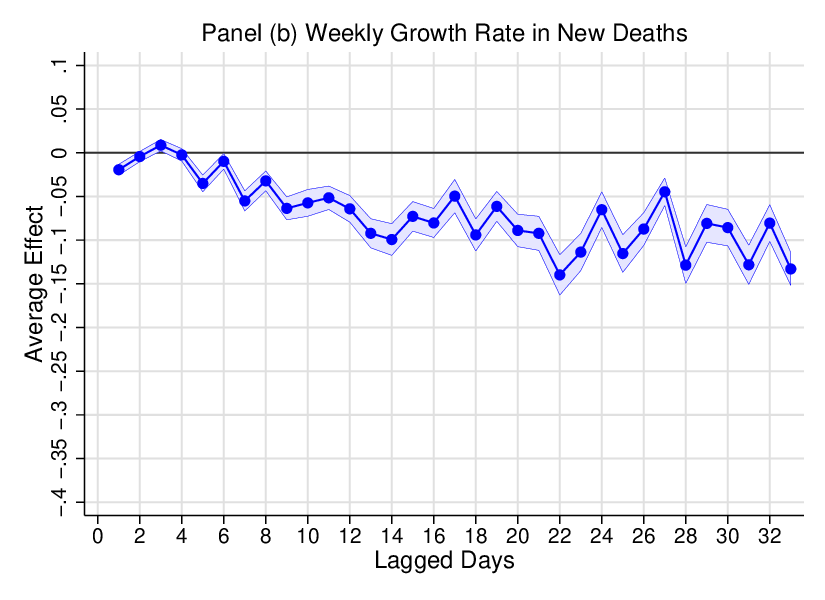}} 
	\caption{\label{figure:att1} Average Treatment Effects of Social Distancing on Growth Rates in New Cases and COVID-Related Deaths.}
\end{figure}

We report the average treatment effects of social distancing in  \Cref{figure:att1}. We find that the social distancing measure effectively reduced the growth rate of infections soon after its introduction, but its effect on deaths became significant after 4-5 days. The social distancing policy's effectiveness was weak initially but improved over time. On average, the social distancing policy reduced the weekly growth rate of confirmed cases by 9.8\% and COVID-related deaths by 7.0\%.

\subsection{Heterogeneous Treatment Effects of Social Distancing}

We then investigate the possible heterogeneity in social distancing effectiveness based on state-level socio-demographic attributes, including income, race, education, political preference, and media consumption. These social variables may influence people's beliefs and compliance behaviors; therefore, the effectiveness of social distancing may vary. We find states with higher income, higher education, more white people, more democratic voters, and higher CNN viewership have a more considerable reduction in the infection growth rate.  

\subsubsection{Income Heterogeneity}

We study the social distancing treatment effects for states with different income levels. High-income households are generally more resilient to economic and health shocks than their low-income counterparts.\footnote{It is important to note that household-level measures do not necessarily translate into aggregated state-level measures.} Additionally, households' responses to the social distancing policy may vary in income. Households with higher incomes are better equipped to cope with working from home and are, therefore, more likely to be compliant with stay-at-home orders. As such, we expect states with higher levels of personal income to experience a more significant reduction in the growth rate of new COVID cases (and deaths) than states with lower personal income.

\begin{figure}[htbp]
	\centering
	\subfloat{\includegraphics[scale=0.55]{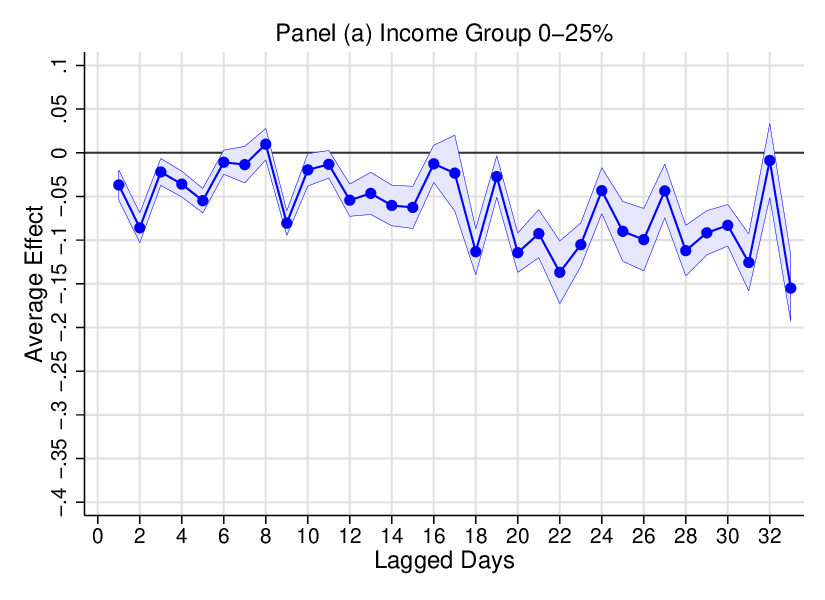}}
	\subfloat{\includegraphics[scale=0.55]{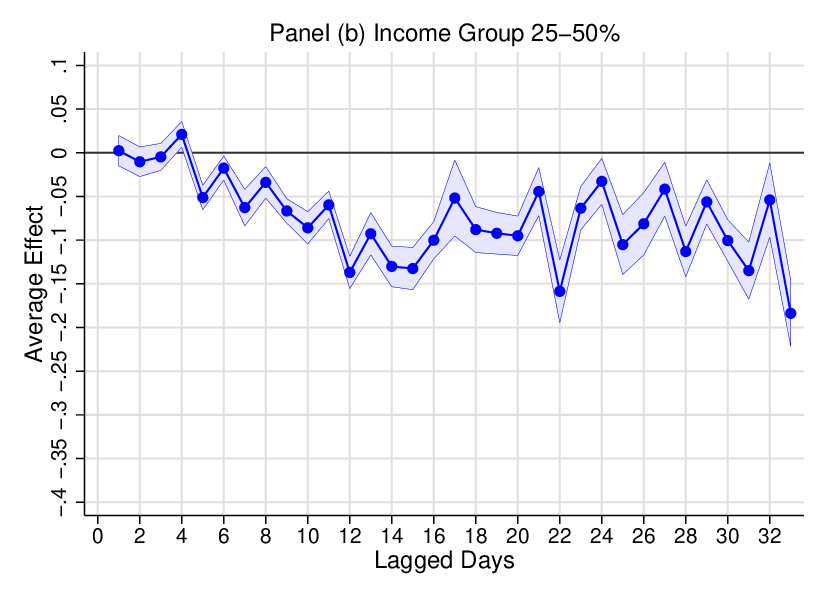}}\\
	\subfloat{\includegraphics[scale=0.55]{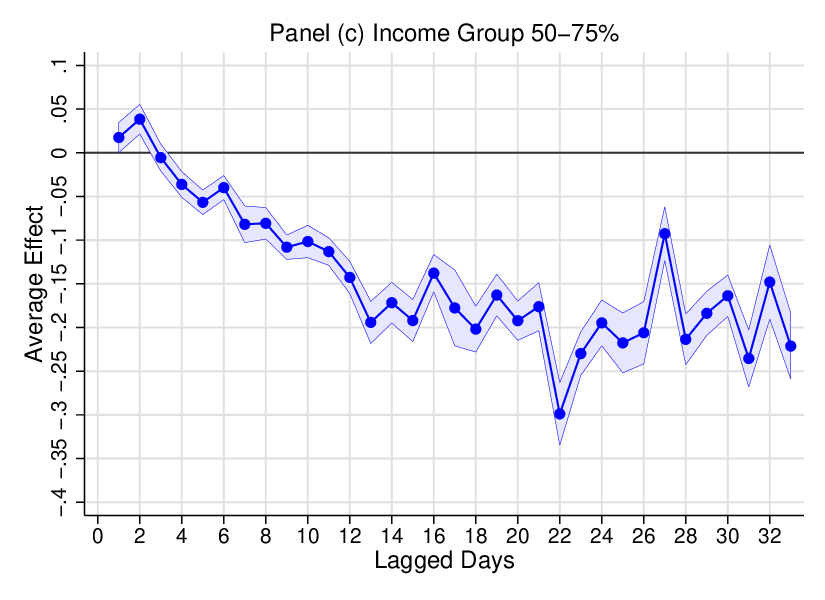}}
	\subfloat{\includegraphics[scale=0.55]{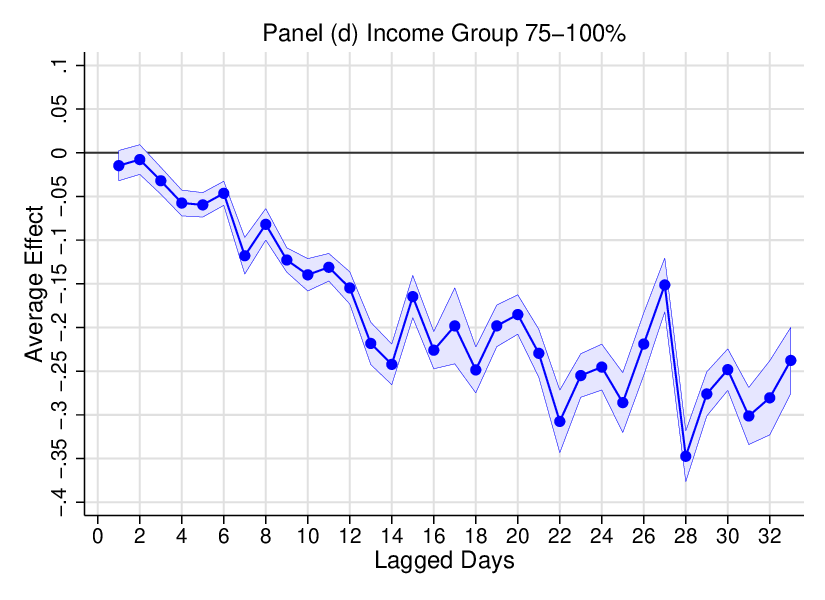}}
	\caption{\label{figure:att_income1} Average Treatment Effects of Social Distancing on Growth Rate in New Cases by Income Group.}
\end{figure}

We use 2019 Personal Income Data from the Bureau of Economic Analysis to identify each state's income level and then divide states into income quartiles. We show the results for the weekly growth rate of new cases in \Cref{figure:att_income1}. There is a clear pattern differentiated by income group.  For the bottom quartile, the social distancing policy is associated with a 3.70\% reduction in weekly growth rate in new cases after 7-days, a 3.78\% reduction after 14-days, a 6.37\% reduction after 21-days, and a 9.00\% reduction after 28-days. The top quartile is associated with a 4.80\% reduction in weekly growth rate in new cases after 7-days, a 15.58\% reduction after 14-days, a 20.71\% reduction after 21-days, and a 25.88\% reduction after 28-days. Overall, the weakest effects are observed for the bottom quartile and the strongest for the top quartile; this difference is persistent for all lagged days we considered.

\subsubsection{Race and Ethnicity Heterogeneity}

In this section, we study the potential heterogeneity in social distancing effects for states with different major races and ethnicities. We obtain the data on race and ethnicity from the American Community Survey of the Census Bureau in 2019. The Bureau estimates the number of residents who identify as being from one or more cultural backgrounds, including First Nations (identified as American Indian or Alaska Native), Asian, Black, Hispanic, Latino, Native Hawaiian or Other Pacific Islander, and White. For simplicity, we focus on the census categories with ``Black alone'' and ``White alone'' and define a state as having a specific cultural makeup if more than 30\% of the state population belongs to this race.

\begin{figure}[htbp]
\begin{center}
	\subfloat{\includegraphics[scale=0.55]{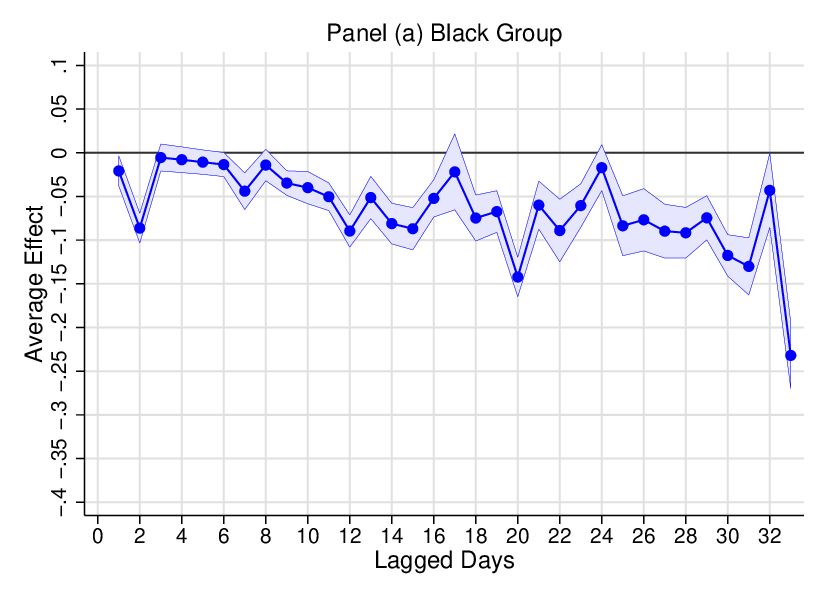}}
	\subfloat{\includegraphics[scale=0.55]{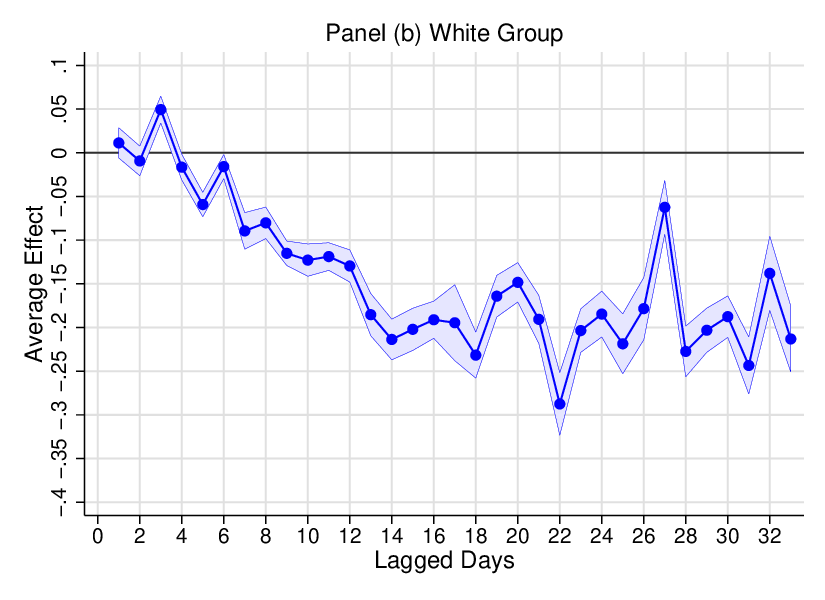}}
	\caption{\label{figure:att_race1} Average Treatment Effects of Social Distancing on  Growth Rate in New Cases by Race.}
\end{center}
\end{figure}

We show the results for the weekly growth rate of new cases in \Cref{figure:att_race1}. There is a large and persistent difference in the social distancing policy effects for states with different cultural profiles. In states where people who identify as Black form the majority of the population, the social distancing policy is associated with a 2.70\% reduction in weekly growth rate in new cases after 7-days, a 5.15\% reduction after 14-days, a 7.22\% reduction after 21-days, and a 7.26\% reduction after 28-days. In states where people who identify as White form the majority of the population, the social distancing policy is associated with a 1.85\% reduction in weekly growth rate in new cases after 7-days, a 13.80\% reduction after 14-days, an 18.90\% reduction after 21-days, and a 19.47\% reduction after 28-days. These results show cultural profile of states significantly affects outcomes from social distancing policies, and different policy solutions may be required for different communities.

\subsubsection{Education Heterogeneity}

In this section, we study the possible heterogeneity in the social distancing effects for states with different education levels. Following \cite{NBERw27560}, we calculate the education level for each state based on the college share within this state and partition all states into four education quartiles.\footnote{In \cite{NBERw27560}, the education level in a state is defined using the PlaceIQ data with the relative share of education in this state in which device $i$ has its permanent residence.}

\begin{figure}[htbp]
\begin{center}
	\subfloat{\includegraphics[scale=0.55]{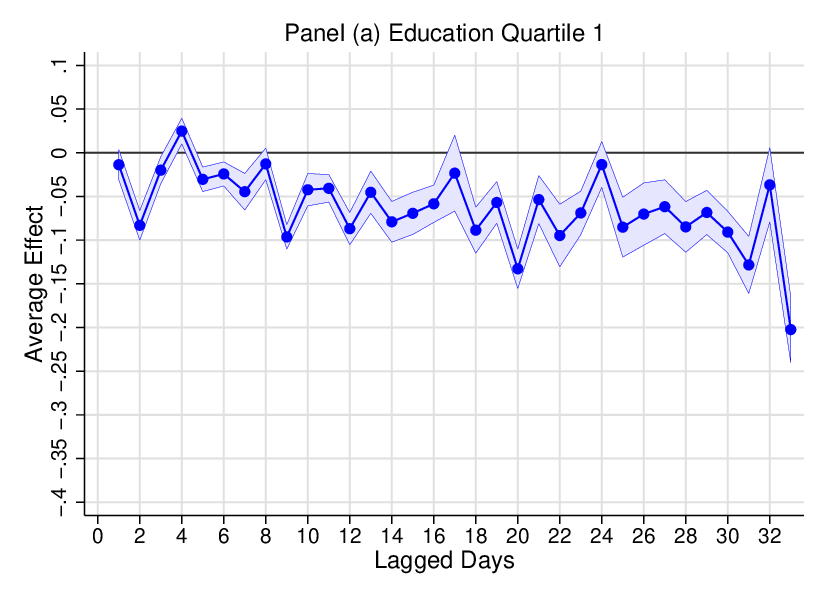}}
	\subfloat{\includegraphics[scale=0.55]{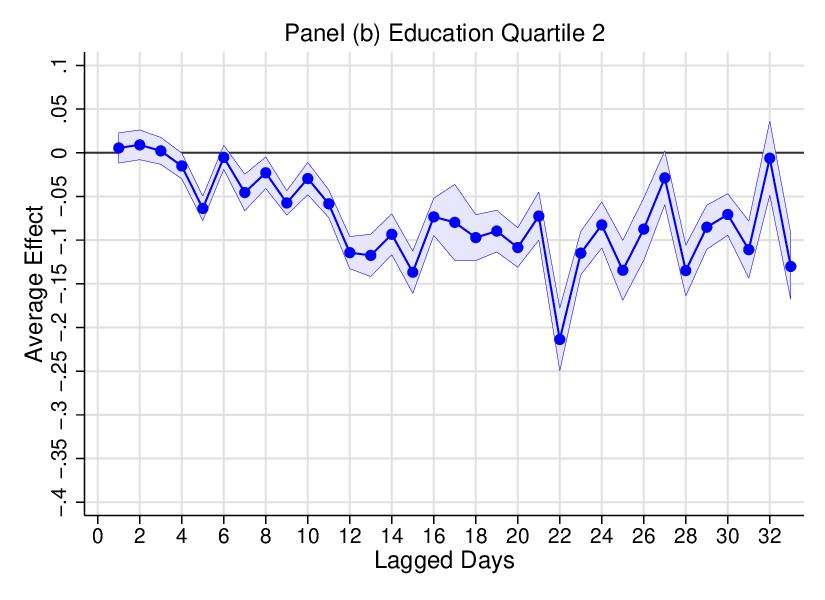}}\\
	\subfloat{\includegraphics[scale=0.55]{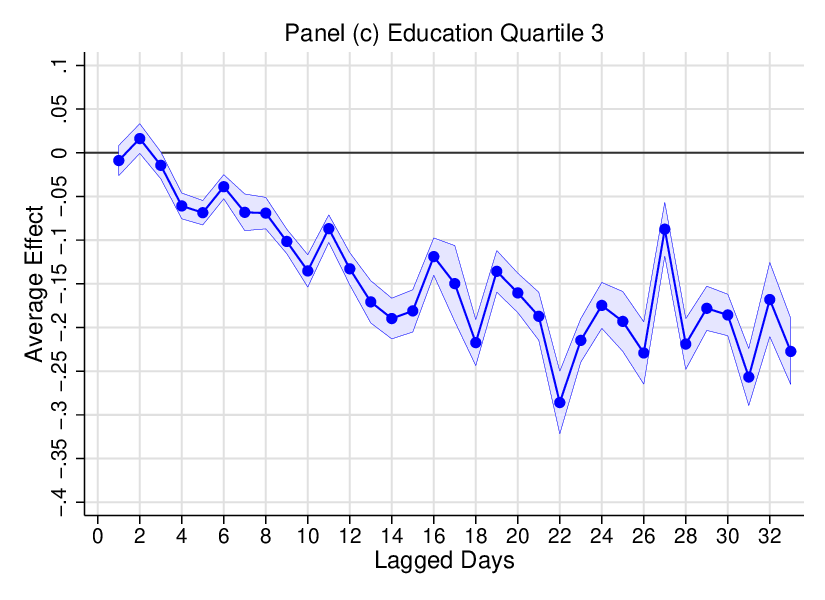}}
	\subfloat{\includegraphics[scale=0.55]{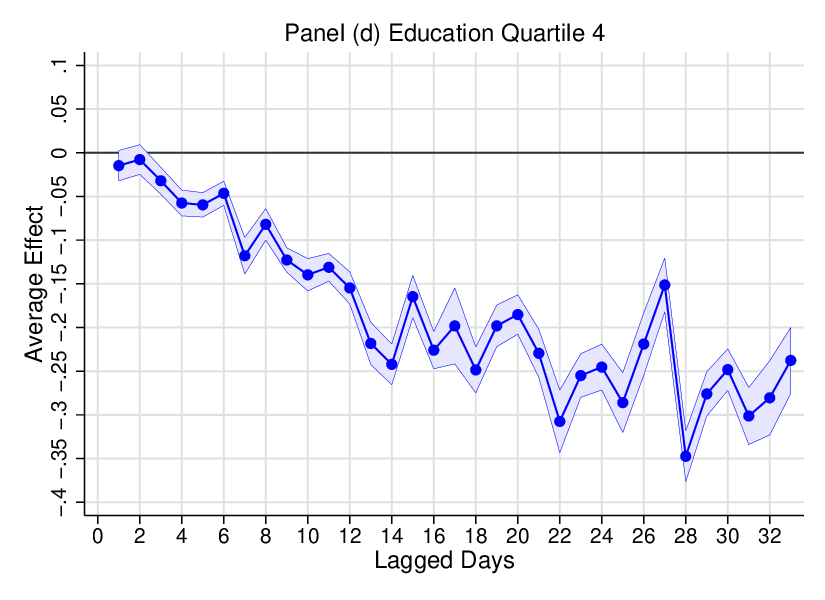}}
	\caption{\label{figure:att_education1} Average Treatment Effects of Social Distancing on  Growth Rate in New Cases by Education.}
\end{center}
\end{figure}

We show the results for the weekly growth rate of new cases in \Cref{figure:att_education1}. We find that the difference in the effects of social distancing by education quartiles is large. For the state in the lowest education quartile, the average reduction in weekly growth rate in new cases following the announcement of the stay-at-home measure is about one-third of the reduction observed in states with the highest level of education. For the bottom education quartile, the social distancing policy is associated with a 2.73\% reduction in the weekly growth rate in new cases after 7-days, a 5.76\% reduction after 14-days, a 6.90\% reduction after 21-days, and a  6.84\% reduction after 28-days. For the top education quartile, it is associated with a 4.79\% reduction in weekly growth rate in cases after seven days, a 15.58\% reduction after 14-days, a 20.71\% reduction after 21-days, and a 25.88\% reduction after 28-days. Overall, states with different education levels range from the weakest for the bottom quartile to the strongest for the top quartile.

\subsubsection{Political Heterogeneity}

States with more Trump supporters are less likely to enact social distancing policies \citep{Adolph2020}. Such political affiliations are also likely to significantly impact personal behavior in response to COVID information and health policies, including day-to-day behaviors and health outcomes.

\begin{figure}[htbp] 
\begin{center}
	\subfloat{\includegraphics[scale=0.55]{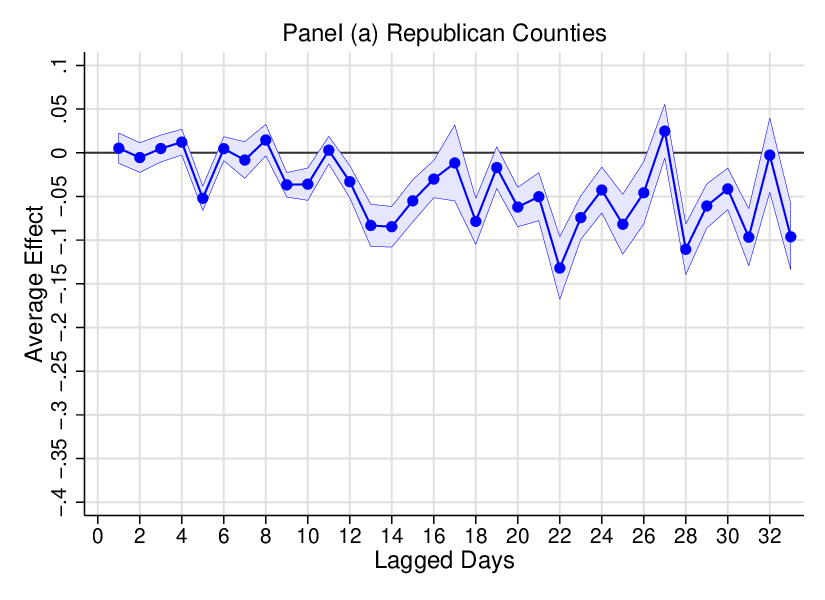}}
	\subfloat{\includegraphics[scale=0.55]{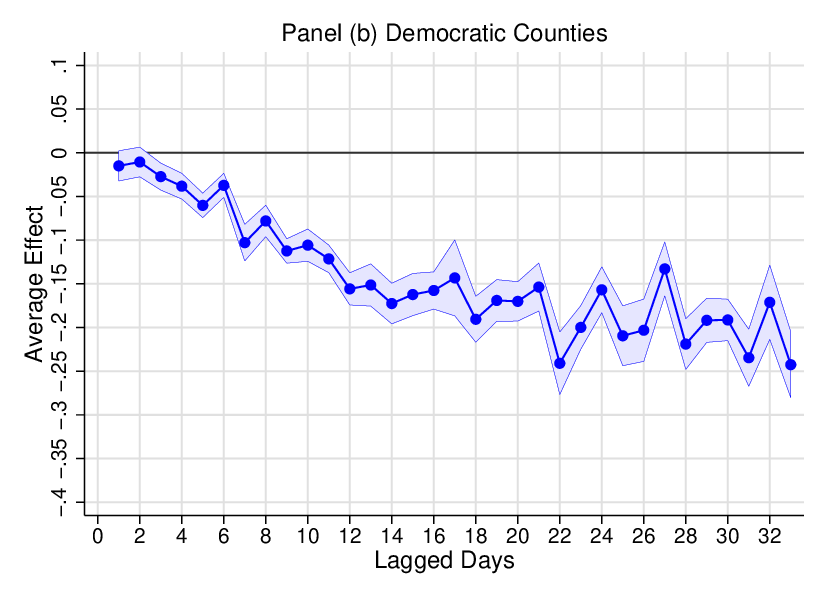}}
	\caption{\label{figure:att_political1} Average Treatment Effects of Social Distancing on  Growth Rate of New Cases by Political Affiliation.}
\end{center}
\end{figure}

In this section, we investigate political heterogeneity in the social distancing treatment effects and use the state-level presidential election returns in 2020 from MIT Election Lab to measure political heterogeneity for each state. We show the results for the weekly growth rate of new cases in \Cref{figure:att_political1}. The social distancing policy is associated with a larger reduction in the growth rate in new cases for democratic states than Republican ones. For the republican states, the social distancing policy is associated with a 4.22\% reduction in the weekly growth rate in new cases after 7-days, a 4.65\% reduction after 14-days, a 5.44\% reduction after 21-days, and a 5.59\% reduction after 28-days. For the democratic states, the social distancing policy is associated with a 5.06\% reduction in the weekly growth rate in cases after 7-days, a 14.02\% reduction after 14-days, a 17.50\% reduction in weekly growth rate in cases after 21 days, and an 18.76\%  reduction in weekly growth rate in cases after 28 days.

\subsubsection{Media Consumption Heterogeneity}		

In this section, we study the effects of social distancing for states with different media news consumption. Studies have demonstrated the persuasive effect of news media on economics and health. An increased Fox News consumption can be associated with a reduced likelihood of staying at home during the pandemic. Therefore, we anticipate that higher consumption of Fox News would lead to lower social distancing effects on reducing the infection growth.

\begin{figure}[htbp] 
\begin{center}
	\subfloat{\includegraphics[scale=0.55]{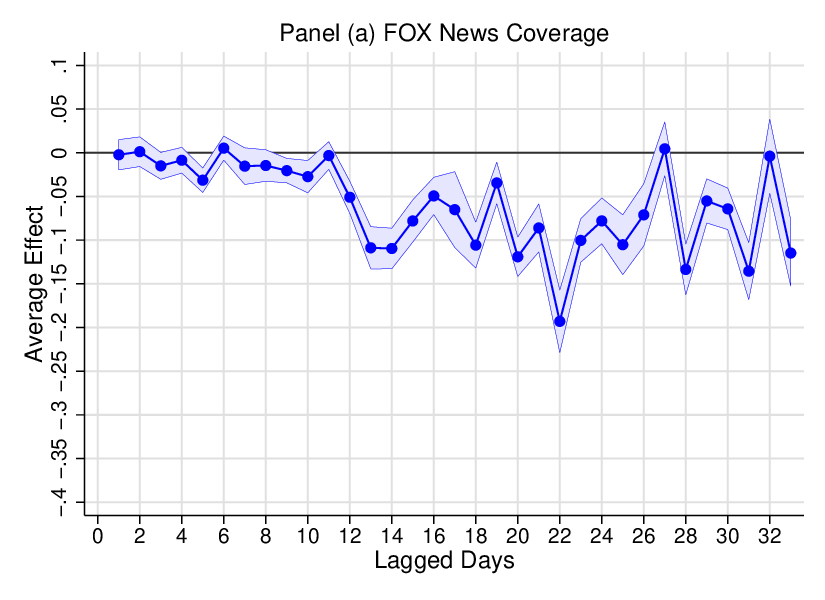}}
	\subfloat{\includegraphics[scale=0.55]{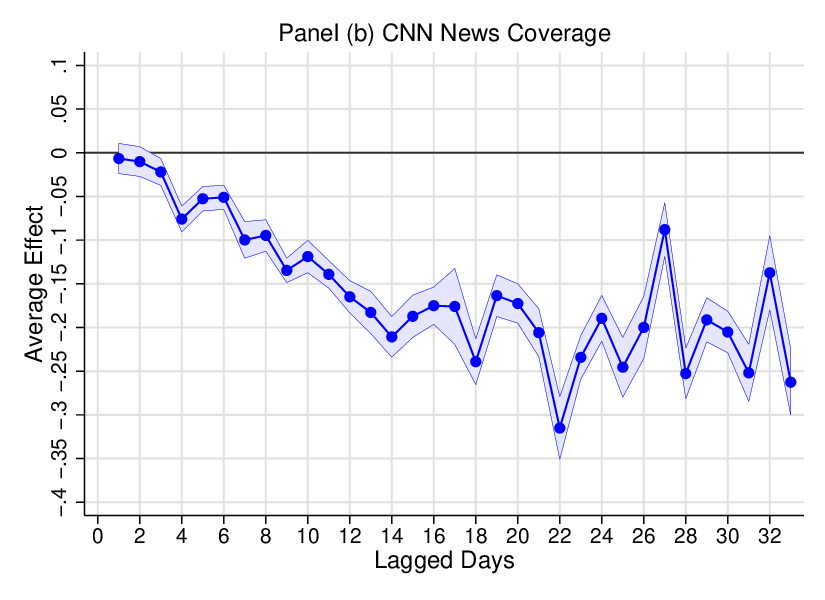}}
	\caption{\label{figure:att_fox1} Average Treatment Effects of Social Distancing on Growth Rate of New Cases by News Consumption.}
\end{center}
\end{figure}

We use the data of Nielsen's monthly NLTV in 2015, which covered 30,517 postal (zip) codes across the US. The data recorded the geographic details and channel positions for a representative sample of all households in the US. We aggregate each household in the data based on state and calculate Fox news consumption across each state. We show the results for the weekly growth rate of new cases in \Cref{figure:att_fox1}. The effects of social distancing on COVID cases and deaths are relatively smaller for states with more Fox news consumption than those states with more CNN news consumption. For states with more CNN news consumption, the social distancing policy is associated with a 4.54\% reduction in the weekly growth rate in new cases after 7-days, a 14.94\% reduction after 14-days, an 18.85\% reduction after 21-days, and a  21.78\% reduction after 28-days. For states with higher levels of FOX news consumption, the social distancing policy is associated with a 0.94\% reduction in the weekly growth rate in new cases after 7-days, a 4.78\% reduction after 14-days, a 7.68\% reduction after 21-days, and a 9.67\% reduction after 28-days.

\section{Discussion}\label{section:discuss}


In this paper, we use a factor imputation methodology \citep{Bai2021} to identify the treatment effects of social distancing policies, specifically stay-at-home orders, across states in the US on reducing the weekly growth rate in new cases and COVID-related deaths. We find social distancing policies reduce the growth rate of new cases and COVID-related deaths; social distancing policies effectively reduce virus transmission. Furthermore, these policies' effectiveness varies over states with different socio-demographic factors and news media consumption.  We complement the literature on evaluating disease control measures and highlight the importance of demographic heterogeneity in affecting these policies' efficacy.


Beyond our findings, we present a useful model to estimate the causal effects of disease control policies and their impact on the spread of infectious diseases. Our model allows us to explicitly specify the counterfactual outcomes for the combinations of treated unit and period under minimal assumptions; this can be useful to analyze the health and economic impacts  (including labor market outcomes) of policy interventions. While establishing the causality relationship between the communities' characteristics and underlying mechanisms is primarily constrained by data availability, we take the first step by providing statistical support on social distancing efficacy and suggest that efforts to `flatten the curve' of transmission rates indeed reduce the virus growth rate both spatially and temporally.

\clearpage
{
\singlespacing
\bibliographystyle{jf}
\bibliography{ref_covid}
}

\clearpage
\appendix
\doublespacing

\begin{center}
{\bf \Large Internet Appendix}
\end{center}

\pagenumbering{arabic}		
\renewcommand*{\thepage}{Appendix-\arabic{page}}

\counterwithin*{table}{section}
\counterwithin*{figure}{section}
\counterwithin*{equation}{section}
\setcounter{table}{0}
\setcounter{figure}{0}
\setcounter{equation}{0}
\renewcommand\thesection{\Alph{section}}
\renewcommand{\thetable}{\Alph{section}.\arabic{table}}
\renewcommand{\thefigure}{\Alph{section}.\arabic{figure}}
\renewcommand{\theequation}{\Alph{section}.\arabic{equation}}

\section{Model}\label{section:appendix_detailed model}

\subsection{Assumptions}\label{section:appendix_assumption}

We denote some quantities before our assumption. For $i = 1,\dots,N$, $t = 1,\dots, T$, $\delta_{N_{0}, N_{1}}=\min \{\sqrt{N_{0}}, \sqrt{N_{1}}\}$, and $K = \dim(\beta)$,
\begin{align*}
\mathbb{V}_{\theta, t}&=\frac{\delta_{N_{0}, N_{1}}^{2}}{N_{0}} \bar{\Lambda}_{\mathcal{T}}^{\prime}\left(\frac{\Lambda^{\prime} \Lambda}{N}\right)^{-1} \Gamma_{t}\left(\frac{\Lambda^{\prime} \Lambda}{N}\right)^{-1} \bar{\Lambda}_{\mathcal{T}}+\frac{\delta_{N_{0}, N_{1}}^{2}}{N_{1}} \sigma_{e}^{2}.\\
\widehat{\mathbb{V}}_{\theta, t} &= \frac{\delta_{N_{0}, N_{1}}^{2}}{N_{0}} \bar{\Lambda}_{\mathcal{T}}^{\prime}\left(\frac{\widehat{\Lambda}^{\prime} \widehat{\Lambda}}{N}\right)^{-1} \Gamma_{t}\left(\frac{\widehat{\Lambda}^{\prime} \widehat{\Lambda}}{N}\right)^{-1} \bar{\Lambda}_{\mathcal{T}}+\frac{\delta_{N_{0}, N_{1}}^{2}}{N_{1}} \widehat{\sigma}_{e}^{2}.\\
\mathbb{V}_{\theta, j, t} &= \frac{\delta_{N_{0}, N_{1}}^{2}}{T_{0}} F_{t}^{\prime}\left(\frac{F^{\prime} F}{T}\right)^{-1} \Phi_{j}\left(\frac{F^{\prime} F}{T}\right)^{-1} F_{t} + \frac{\delta_{N_{0}, N_{1}}^{2}}{N_{0}} \bar{\Lambda}_{j}^{\prime}\left(\frac{\Lambda^{\prime} \Lambda}{N}\right)^{-1} \Gamma_{t}\left(\frac{\Lambda^{\prime} \Lambda}{N}\right)^{-1} \bar{\Lambda}_{j}
\\
&+\delta_{N_{0}, N_{1}}^{2} \sigma_{e,t}^{2}.\\
\widehat{\mathbb{V}}_{\theta, j, t} &= \frac{\delta_{N_{0}, N_{1}}^{2}}{T_{0}} F_{t}^{\prime}\left(\frac{F^{\prime} F}{T}\right)^{-1} \Phi_{j}\left(\frac{F^{\prime} F}{T}\right)^{-1} F_{t}  + \frac{\delta_{N_{0}, N_{1}}^{2}}{N_{0}} \bar{\Lambda}_{j}^{\prime}\left(\frac{\widehat{\Lambda}^{\prime} \widehat{\Lambda}}{N}\right)^{-1} \Gamma_{t}\left(\frac{\widehat{\Lambda}^{\prime} \widehat{\Lambda}}{N}\right)^{-1} \bar{\Lambda}_{j}
\\
&+\delta_{N_{0}, N_{1}}^{2} \widehat{\sigma}_{e,t}^{2}.\\
\widehat{\sigma}_{e}^{2} &=\frac{1}{T N_{0}-r\left(T+N_{0}\right)+r^{2}-K} \sum_{i \leq N_{0}} \sum_{t=1}^{T} \widehat{e}_{i t}^{2}.\\
\widehat{\sigma}_{e,t}^{2} &= \frac{1}{N-1} \sum_{i \neq j}^{N-1} \widehat{e}_{i t}^{2}.
\end{align*}

\begin{assumption}\label{assumption:1}
There exists a constant $M$ such that
\begin{enumerate} 
	\item (Factors)
	\begin{enumerate}
		\item $E\left\|F_{t}^{0}\right\|^{4} \leq M \leq \infty$,
		\item $T^{-1}\sum_{t=1}^{T}F_{t}^{0} F_{t}^{0\prime} \stackrel{p}{\longrightarrow} \Sigma_{F}$, where $\Sigma_{F}$ is a positive definite matrix.
	\end{enumerate}	
	\item (Factor Loadings)
	\begin{enumerate}
		\item $\left\|\Lambda_{i}^{0}\right\| \leq M$,
		\item $N^{-1}\sum_{i=1}^{N} \Lambda_{i}^{0\prime} \Lambda_{i}^{0} \stackrel{p}{\longrightarrow} \Sigma_{\Lambda}$, where $\Sigma_{\Lambda}$ is a positive definite matrix,
		\item The eigenvalues of $\Sigma_{F} \Sigma_{\Lambda}$ are distinct.
	\end{enumerate}
	\item (Error terms):
	\begin{enumerate}
		\item $e_{it} \sim \mathcal{N}(0,\sigma_{e,t}^2)$, $\forall i$,
		\item $E\left(N^{-1}\sum_{i=1}^{N} e_{i t} e_{i s}\right)=\gamma_{N}(s, t)$, where $\sum_{t=1}^{T}\left|\gamma_{N}(s, t)\right| \leq M, \forall s$,
		\item $E\left(e_{i t} e_{j t}\right)=\tau_{i j, t}$, where $\left|\tau_{i j, t}\right| \leq\left|\tau_{i j}\right|$ for some $\tau_{i j}$, and $\sum_{j=1}^{N}\left|\tau_{i j}\right| \leq M, \forall i$,
		\item $E\left(e_{i t} e_{j s}\right)=\tau_{i j, s t}$ and $(NT)^{-1} \sum_{i=1}^{N} \sum_{j=1}^{N} \sum_{t=1}^{T} \sum_{s=1}^{T}\left|\tau_{i j, t s}\right|<M$,
		\item For every $(t, s)$, $E \left(N^{-1 / 2} \sum_{i=1}^{N}\left[e_{i s} e_{i t}-E\left(e_{i s} e_{i t}\right)\right]^{4} \right)\leq M$,
		\item $E\left(N^{-1} \sum_{i=1}^{N}\left\|\frac{1}{\sqrt{T}} \sum_{t=1}^{T} F_{t}^{0} e_{i t}\right\|^{2}\right) \leq M$.
	\end{enumerate}
	\item (Central Limit Theorems)
	\begin{enumerate}
		\item For each $i$, as $N \rightarrow \infty$, $N^{-1 / 2} \sum_{i=1}^{N} \Lambda_{i}^{0} e_{i t} \stackrel{d}{\longrightarrow} N\left(0, \Gamma_{t}\right)$, 
		\item For each $t$,  as $T \rightarrow \infty$, $T^{-1 / 2} \sum_{t=1}^{T} F_{t}^{0} e_{i t} \stackrel{d}{\longrightarrow} N\left(0, \Phi_{i}\right)$.
	\end{enumerate}
\end{enumerate}
\end{assumption}

\begin{assumption}\label{assumption:2}
Order conditions:
\begin{enumerate}
	\item The conditions $T\times N_{0}> r(T +N_{0})$ and $N\times T_{0}> r(N +T_{0})$ are satisfied for any $N, T, N_{0}, T_{0}$,
	\item $\frac{\sqrt{N}}{\min \left\{N_{0}, T_{0}\right\}} \rightarrow 0$
	as $N_{0}, N \rightarrow \infty$ and $T_{0}, T \rightarrow \infty$,
	\item $\frac{\sqrt{T}}{\min \left\{N_{0}, T_{0}\right\}} \rightarrow 0$ as $N_{0}, N \rightarrow \infty$ and $T_{0}, T \rightarrow \infty$.
\end{enumerate}
\end{assumption}

\begin{assumption}\label{assumption:3}
Factors and factor loadings:
\begin{enumerate}
	\item $\frac{\Lambda_{0}^{0 \prime} \Lambda_{0}^{0}}{N_{0}} \stackrel{p}{\longrightarrow} \Sigma_{\Lambda, 0}>0, \frac{\Lambda_{m}^{0 \prime} \Lambda_{m}^{0}}{N_{m}} \stackrel{p}{\longrightarrow} \Sigma_{\Lambda, m}>0, \frac{1}{\sqrt{N_{0}}} \sum_{i=1}^{N_{0}} \Lambda_{i}^{0} e_{i t} \stackrel{d}{\longrightarrow} N\left(0, \Gamma_{0 t}\right)$,
	\item $\frac{F_{0}^{0 \prime} F_{0}^{0}}{T_{0}} \stackrel{p}{\longrightarrow} \Sigma_{F, 0}>0, \frac{F_{m}^{0 \prime} F_{m}^{0}}{T_{m}} \stackrel{p}{\longrightarrow} \Sigma_{F, m}>0, \frac{1}{\sqrt{T_{0}}} \sum_{s=1}^{T_{0}} F_{s}^{0} e_{i s} \stackrel{d}{\longrightarrow} N\left(0, \Phi_{0 i}\right)$.
\end{enumerate}
\end{assumption}

Part (1) and (2) of \Cref{assumption:1} impose identification conditions on the factor structure and the low-rank component. Part (3) of \Cref{assumption:1} assumes the error terms to be weakly correlated in cross-sectional and time-series dimensions. Part (4) of \Cref{assumption:1} assumes the large sample distribution of factor estimates. \Cref{assumption:2} imposes that the order condition holds for both TALL and WIDE block matrices. \Cref{assumption:3} requires the subsample matrix's identification condition such that the subsample moment matrices for the factor and factor loadings to be positive definite.

We further explain the common factors may capture differential time trends with the following example. Consider the special factor and factor loading in the form of
\begin{align}
	F_t = \left[ \begin{array}{c}
		1   \\
		\xi_t
	\end{array} \right] \qquad \mathrm{and} \qquad %
	\Lambda_i = \left[ \begin{array}{c}
		\alpha_i   \\
		1
	\end{array} \right]
	\label{Eq Special Factor and Factor Loadings}
\end{align}
for all $ i $ and $ t $. Then, \Cref{Eq Special Factor and Factor Loadings} can be further rewritten as
\begin{align*}
	Y_{it} %
	&= \theta_{it} \mathcal{D}_{it} + X_{i,t-m}^{'} \beta + Z_{i,t}^{'} \gamma + \Lambda_{i}^{'} F_t + e_{it}   \\
	&= \theta_{it} \mathcal{D}_{it} + X_{i,t-m}^{'} \beta + Z_{i,t}^{'} \gamma + \alpha_i + \xi_t + e_{it},
\end{align*}
where the interactive effect $ \Lambda_{i}^{'} F_t $ becomes the individual effects $ \alpha_i $ and the time effects $ \xi_t $ entering our model additively. Consequently, we believe that our common factors effectively capture the differential time trends across states.

\subsection{Estimation}\label{section:appendix_estimation}

In this section, we present the computation algorithm to estimate both average treatment effects and individual treatment effects that allow the pre-treatment period for each state in the treated group to be different.


\vspace{0.5cm}
\noindent 
\textbf{Algorithm.}
We introduce the following algorithm to calculate the average treatment effects and individual treatment effects:
\begin{enumerate}
\item We estimate the coefficient of $\beta$ and $\gamma$ using the observations in the control group with interactive fixed effects method \citep{Bai2009}. We then estimate the residual $R_{it} $ for the whole observations where $R_{it} \equiv Y_{it} - X_{i,t-14}'\widehat{\beta} - Z_{it}'\widehat{\gamma} $.
\item We estimate the latent factor $F$ using the observations in the $\textsc{Tall}$ block matrix of residual $R_{\textsc{Tall}}$ with asymptotic principal components method.
\item We estimate the factor loadings $\Lambda$ using two-step approach:
\begin{enumerate}
	\item For $i = 1,\dots, N$, create the new matrix $\widetilde{F}_{i} = [\mathbb{I}_{i}~F_{i}]$, where $F_{i}$ is the submatrix of $F$ with the row from $1:T_{0,i}$ and $\mathbb{I}_{i}$ is $T_{0,i} \times 1$ matrix of one.
	\item For $i = 1,\dots,N$, estimate the $\widetilde{\Lambda}_{i} = (\widetilde{F}_{i}'\widetilde{F}_{i})^{-1}\widetilde{F}_{i}'R_{it}$.
\end{enumerate}
\item We estimate the missing value with $\widehat{Y}_{i t}(0)=X_{i t}^{\prime} \widehat{\beta}+\widetilde{C}_{i t}$, where $\widetilde{C} = [\mathbb{I}_{T}~\widetilde{F}] \times \widetilde{\Lambda}'$, $\mathbb{I}_{T}$ is $T \times 1$ matrix of one.
\item We compute the individual treatment effects $\widehat{\theta}_{i t} = Y_{it}(1) - \widehat{Y}_{i t}(0),~i \in \mathcal{T},~t =T_{0,i}+1, \ldots, T$.
\item We compute the average treatment effects on treated $\widehat{\theta}_{t}=\frac{1}{N_{1}} \sum_{i \in \mathcal{T}} \widehat{\theta}_{i t}$.
\item We compute the average treatment effects on treated for the individuals in the group $j$, $\widehat{\theta}_{j,t}=\frac{1}{N_{j}} \sum_{i \in \mathcal{J}} \widehat{\theta}_{i t}$, where $i \in \mathcal{J}$, and $N_{j}$ is the sample size of the group $j$.
\end{enumerate}

Note that in Steps 2 and 3, we estimate both latent factors $F$ and factor loadings $\Lambda$ for both TALL block and WIDE block with asymptotic principal components method and the number of factors for these two blocks $r_{\text {TALL}}$ and $r_{\text{WIDE}}$ can be consistently estimated using the method developed in \cite{bai2002determining,bai2019rank}. When the number of factors of the two blocks does not coincide, we choose the number $r=\max \left(r_{\text {TALL}}, r_{\text {WIDE}}\right)$ and use $r$ for the estimation in Steps 2 and 3. Therefore, we can treat the number of factors known for the TALL and WIDE blocks in the estimation procedure.


\subsection{Large Sample Properties}\label{section:appendix_large sample}

In this section, we establish the asymptotic distribution results for the average treatment effects on treated $\widehat{\theta}_{t}$ and the heterogenous treatment effects $\widehat{\theta}_{j,t}$ for a group of $j$ that belongs to the treated group $\mathcal{T}$ under the assumptions in \Cref{section:appendix_assumption}.

\begin{proposition}\label{proposition:1}
Let \Cref{assumption:1}, \Cref{assumption:2}, and \Cref{assumption:3} hold. As $N_{0}, T_{0}, N_{1} \rightarrow \infty$, we have:
\begin{enumerate}
	\item The average treatment effects on treated $\widehat{\theta}_{t}$ is asymptotically normal:
	\[
	\delta_{N_{0}, N_{1}}\left(\frac{\widehat{\theta}_{t}-\theta_{t}}{\sqrt{\mathbb{V}_{\theta, t}}}\right) \stackrel{d}{\longrightarrow} N(0,1).
	\]
	\item The asymptotic variance $\widehat{\sigma}_{e,t}^2$ can be consistently estimated by: 
	\[
	\widehat{\sigma}_{e,t}^{2}= \widehat{\mathbb{V}}_{\theta, t} / \delta_{N_{0}, N_{1}}.
	\]
\end{enumerate}
\end{proposition}

\begin{proposition}\label{proposition:2}
Let \Cref{assumption:1}, \Cref{assumption:2}, and \Cref{assumption:3} hold. As $N_{0}, T_{0}, N_{1}, N_{j} \rightarrow \infty$, we have:
\begin{enumerate}
	\item The heterogeneous treatment effects on group $j$, $\widehat{\theta}_{j}$ is asymptotically normal:
	\begin{align*}
	\delta_{N_{0}, N_{j}}\left(\frac{\widehat{\theta}_{j}-\theta_{j}}{\sqrt{\mathbb{V}_{\theta, j}}}\right) \stackrel{d}{\longrightarrow} N(0,1).
	\end{align*}
    where $ \theta_{j} = \frac{1}{T_1} \sum_{s>T_0} \theta_{js} $...

	\item The asymptotic variance $\widehat{\sigma}_{j}^{2}$ can be consistently estimated by: 
	\[
	\widehat{\sigma}_{j}^{2}
	= \widehat{\mathbb{V}}_{\theta, j} / \delta_{N_{0}, N_{j}}.
	\]
\end{enumerate}
\end{proposition}

Our estimation procedure provides a robust and reliable estimate and inference for treatment effects without relying on restrictive distributional assumptions for error terms or parallel trend assumptions for average outcomes of both treated and control units in the absence of treatment. 

In the context of social distancing policies, which are endogenously enacted by health authorities, recent research has employed either a difference-in-differences approach or an instrumental variable approach to address endogeneity concerns. While these studies confirm the effectiveness of social distancing, they may overlook crucial observed and unobserved factors that can confound the relationship. Specifically, there exists a reverse causality between social distancing and the growth in new cases, as state governments introduce restrictions in response to increases in cases and deaths. Consequently, variations in policies may reflect local case numbers rather than the true causal effect of the policy. Additionally, the voluntary precaution effect, whereby people take precautions before government restrictions are announced in response to publicized COVID outbreaks, can lead to biased difference-in-difference estimates that find a spurious negative relationship between social distancing requirements and growth in new cases. Furthermore, the anticipation effect of governments announcing policies ahead of time may change residents' behavior in response to the policy information itself. Lastly, spillover effects from other states can affect the effectiveness of one state's social distancing measures on the growth of new cases in another state, biasing the difference-in-difference estimates towards zero.

In contrast, our factor model approach consistently estimates the average and heterogeneous treatment effects by identifying the causal relationship between social distancing and reductions in COVID-19 infections while accounting for factors that influence virus transmission over time and across states. Our method can trace the policy's effect on each unit before and after it takes effect, effectively alleviating the endogeneity concerns discussed above. By considering pre-treatment and post-treatment outcomes, we capture the potential reverse causality effect on outcomes in the days leading to the policy and the voluntary precaution effect that improves outcomes for treated units before the policy implementation. Moreover, our approach considers the difference between treatment and control groups using different dimensions by controlling for both observed confounders in the regression and interactive fixed effects that remove the partial correlation that may vary across units and over time \citep{Bai2021}. This comprehensive approach ensures that our estimates are unbiased and accurately reflect the true causal impact of social distancing policies on the spread of COVID-19.

\clearpage
\section{Proofs}\label{section:appendix_proof}	

\subsection{Proofs of \Cref{proposition:1}}

\noindent


The average treatment effects on treated is defined as:
\begin{equation}\label{equation:att1}
\theta_{t}=\frac{1}{N_{1}}\left[\sum_{i \in \mathcal{T}} Y_{i t}(1)-Y_{i t}(0)\right],
\end{equation}

It follows that
\begin{align}
\label{equation:att2}
\begin{split}
	Y_{i t}(1)-\widehat{Y}_{i t}(0)&=\theta_{it}+X_{i t}^{\prime}(\beta-\widehat{\beta})+Z_{i t}^{\prime} \left( \gamma - \widehat{\gamma} \right)+C_{i t}-\widehat{C}_{i t}+e_{i t},\\
	Y_{i t}(0)-\widehat{Y}_{i t}(0)&=X_{i t}^{\prime}(\beta-\widehat{\beta})+Z_{i t}^{\prime} \left( \gamma - \widehat{\gamma} \right)+C_{i t}-\widehat{C}_{i t}+e_{i t}.
\end{split}
\end{align}

Combining \Cref{equation:att1} and \Cref{equation:att2}, we have
\begin{align}
	\widehat{\theta}_{t} - \theta_{t} %
	&= \frac{1}{N_{1}} \sum_{i \in \mathcal{T}} X_{i t}^{\prime} \left( \beta - \widehat{\beta} \right) + \frac{1}{N_{1}} \sum_{i \in \mathcal{T}} Z_{i t}^{\prime} \left( \gamma - \widehat{\gamma} \right) + \frac{1}{N_{1}} \sum_{i \in \mathcal{T}} \left( C_{i t}-\widehat{C}_{i t} \right) + \frac{1}{N_{1}} \sum_{i \in \mathcal{T}} e_{i t} \label{Eq theta_t^hat - theta_t}   \\
	&= A_{1} + A_{2} + A_{3} + A_{4}. \nonumber
\end{align}

By \Cref{assumption:1}, we have:
\[
A_{4} = \frac{1}{N_{1}} \sum_{i \in \mathcal{T}} e_{i t}=O_{p}(\frac{1}{\sqrt{N_{1}}}).
\]

By \Cref{assumption:2}, $\beta$ is homogeneous across $i$ and $t$, we
apply the Theorem 1 in \cite{Bai2009}:
\begin{align*}
	A_1 = \frac{1}{N_{1}} \sum_{i \in \mathcal{T}} X_{i t}^{\prime} \left( \beta - \widehat{\beta} \right) = O_{p} \left(\frac{1}{\sqrt{N_0 T_0}}\right) %
	\quad \mathrm{and} \quad %
	A_2 = \frac{1}{N_{1}} \sum_{i \in \mathcal{T}} Z_{i t}^{\prime} \left( \gamma - \widehat{\gamma} \right) = O_{p} \left(\frac{1}{\sqrt{N_0 T_0}}\right).
\end{align*}

We decompose the term $A_{3}$ as
\[
\begin{aligned}
A_{3} = & \frac{1}{N_{1}} \sum_{i \in \mathcal{T}}\left(\widehat{C}_{i t}-C_{i t}\right)\\= & F_{t}^{\prime}\left(\frac{F^{\prime} F}{T}\right)^{-1} \mathbf{B}_{F} \frac{1}{T_{0} N_{1}}\left(\sum_{i \in \mathcal{T}} \sum_{s=1}^{T_{0,i}} F_{s} e_{i s}\right)+\bar{\Lambda}_{\mathcal{T}}\left(\frac{\Lambda^{\prime} \Lambda}{N}\right)^{-1} \mathbf{B}_{\Lambda} \frac{1}{N_{0}} \sum_{k=1}^{N_{0}} \Lambda_{k} e_{k t}
\\
&+O_{p}\left(\delta_{N_{0}, T_{0}}^{-2}\right)\\ = & A_{31} + A_{32} +A_{33},
\end{aligned}
\]
where is $\mathbf{B}_{F}$ defined as $\frac{T_{o}}{T} I_{r}+\frac{T_{m}}{T}\left(\frac{F_{m}^{0 \prime} F_{m}^{0}}{T_{m}}\right)\left(\frac{F_{o}^{0 \prime} F_{o}^{0}}{T_{o}}\right)^{-1}$ and $\mathbf{B}_{\Lambda}$  is defined as $\frac{N_{o}}{N} I_{r}+\frac{N_{m}}{N}\left(\frac{\Lambda_{m}^{0 \prime} \Lambda_{m}^{0}}{N_{m}}\right)\left(\frac{\Lambda_{o}^{\prime} \Lambda_{o}}{N_{o}}\right)^{-1}$.

By \Cref{assumption:1},
\[
A_{31} =  O_{p}\left(\frac{1}{\sqrt{T_{0} N_{1}}}\right).
\]

We have:
\begin{align}
\widehat{\theta}_{t}-\theta_{t} = %
\bar{\Lambda}_{\mathcal{T}}^{\prime}\left(\frac{\Lambda^{\prime} \Lambda}{N}\right)^{-1} \mathbf{B}_{\Lambda} \frac{1}{N_{0}} \sum_{k=1}^{N_{0}} \Lambda_{k} e_{k t}+\frac{1}{N_{1}} \sum_{i \in \mathcal{T}} e_{i t} %
+O_{p}\left(\frac{1}{\sqrt{ T_{0}N_{0}}}\right)+O_{p}\left(\frac{1}{\sqrt{T_{0} N_{1}}}\right)+O_{p}\left(\delta_{N_{0}, T_{0}}^{-2}\right).
\end{align}

\qed



\subsection{Proofs of \Cref{proposition:2}}

\noindent


The average treatment effects on treated for group $j$ are defined as:

\begin{equation}\label{equation:itt1}
\theta_{j,t}=\frac{1}{N_{j}}\left[\sum_{i \in \mathcal{J}} Y_{i t}(1)-Y_{i t}(0)\right],
\end{equation}

It follows that
\begin{align}\label{equation:itt2}
\begin{split}
	Y_{i t}(1)-\widehat{Y}_{i t}(0)&=\theta_{it}+X_{i t}^{\prime}(\beta-\widehat{\beta})+C_{i t}-\widehat{C}_{i t}+e_{i t},\\
	Y_{i t}(0)-\widehat{Y}_{i t}(0)&=X_{i t}^{\prime}(\beta-\widehat{\beta})+C_{i t}-\widehat{C}_{i t}+e_{i t}.
\end{split}
\end{align}

The proof of \Cref{proposition:2} are similar to that of \Cref{proposition:1}, and hence, we only highlight those different parts. Recall \Cref{equation:itt1} and \Cref{equation:itt2} and consider the estimation of treatment effect on a single unit $ j $:
\begin{align}
	\widehat{\theta}_{jt} - \theta_{jt} %
	&= X_{j t}^{\prime} \left( \beta - \widehat{\beta} \right) + Z_{j t}^{\prime} \left( \gamma - \widehat{\gamma} \right) + \left( C_{j t}-\widehat{C}_{j t} \right) + e_{j t} \label{Eq theta_jt^hat - theta_jt}   \\
	&= X_{j t}^{\prime} \left( \beta - \widehat{\beta} \right) + Z_{j t}^{\prime} \left( \gamma - \widehat{\gamma} \right)   \nonumber \\
	&\quad - F_{t}^{\prime}\left(\frac{F^{\prime} F}{T}\right)^{-1} \mathbf{B}_{F} \frac{1}{T_{0}} \left( \sum_{s=1}^{T_{0,j}} F_{s} e_{j s} \right) - \Lambda_j^\prime \left( \frac{\Lambda^{\prime} \Lambda}{N} \right)^{-1} \mathbf{B}_{\Lambda} \frac{1}{N_{0}} \sum_{k=1}^{N_{0}} \Lambda_{k} e_{k t} + e_{j t} + O_{p} \left(\delta_{N_{0}, T_{0}}^{-2}\right).
\end{align}
Then, it is similar to the proof of \Cref{proposition:1} to verify \Cref{proposition:2}.




%
%
%

\qed

\clearpage
\section{Robustness Results}\label{section:appendix_results}


In this section, we conduct two robustness checks to further validate our findings. The first check uses state-level data on non-essential business closures as a measure of social distancing, while the second extends the time window beyond April 20 to examine whether cases rebounded after the lifting of stay-at-home orders.

For the first robustness test, we create a social distancing policy index $D_{i,t}$ for each state, which equals 1 if non-essential business closures are announced in state $i$ on day $t$ and 0 otherwise. Using non-essential business closures as the measure for social distancing, we have $N_{0}$ = 24, $T_{0}$ = 28, $N$ = 55, and $T$ = 61. Therefore, \Cref{assumption:2}f, which states that $\frac{\sqrt{N}}{\min \left\{N_{0}, T_{0}\right\}} \rightarrow 0$ and $\frac{\sqrt{T}}{\min \left\{N_{0}, T_{0}\right\}} \rightarrow 0$ as $N_{0}, N \rightarrow \infty$ and $T_{0}, T \rightarrow \infty$, is satisfied for the state-level data we considered.

\begin{figure}[htbp]
	\begin{center}
		\subfloat{\includegraphics[scale=0.55]{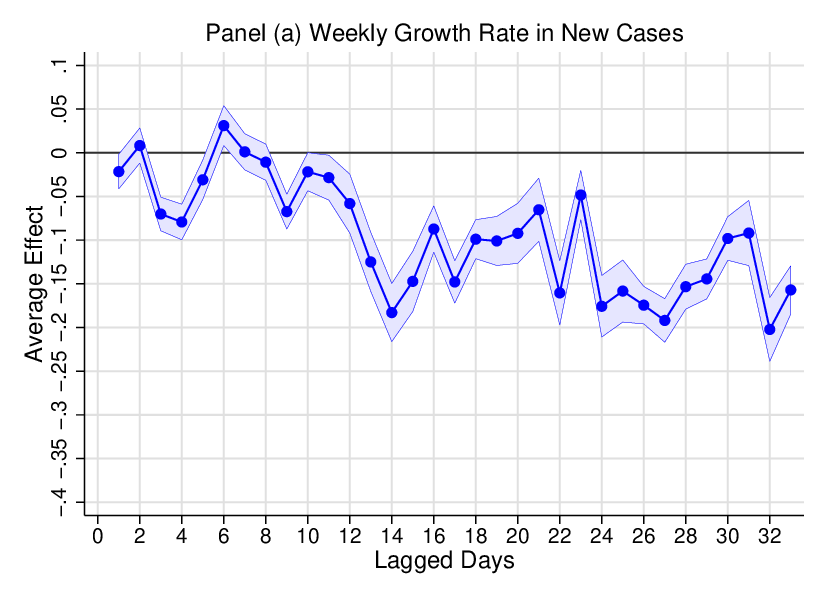}}
		\subfloat{\includegraphics[scale=0.55]{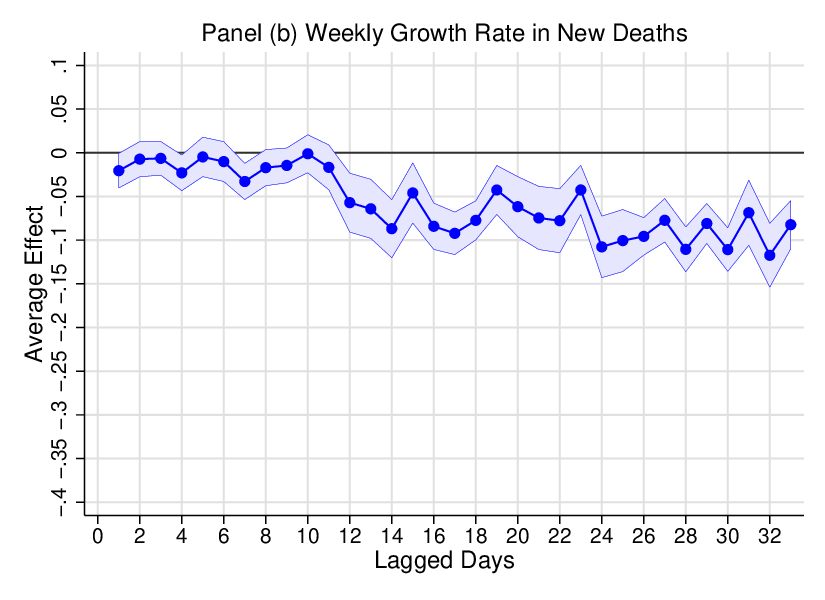}}
		\caption{\label{figure:att2} {Average Treatment Effects of Social Distancing on the Growth Rate of New Infections and COVID-Related Deaths Using Non-Essential Business Closures as the Policy Variable.}}
	\end{center}
\end{figure}

Panel (a) of \Cref{figure:att2} presents the dynamic effect of the state-level social distancing policy on the growth rate of infections with 95\% confidence intervals. The social distancing policy is associated with a 2.30\% reduction in the weekly growth rate of new cases after 7 days, a 7.06\% reduction after 14 days, a 10.57\% reduction after 21 days, and a 15.18\% reduction after 28 days. Panel (b) of \Cref{figure:att2} displays the dynamic effect of state-level social distancing policies on the growth rate of COVID-related deaths with 95\% confidence intervals. The policy starts to reduce the growth rate of COVID-related deaths from the fifth day after its announcement and is associated with a 1.49\% reduction in the growth of weekly COVID-related deaths after 7 days, a 3.68\% reduction after 14 days, a 6.84\% reduction after 21 days, and an 8.74\% reduction after 28 days.

\begin{figure}[htbp]
	\begin{center}
		\subfloat{\includegraphics[scale=0.55]{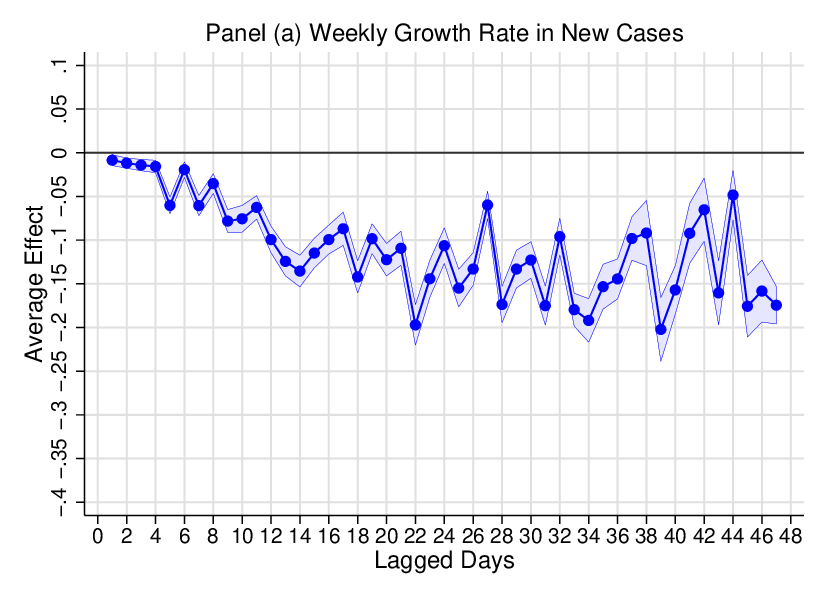}}
		\subfloat{\includegraphics[scale=0.55]{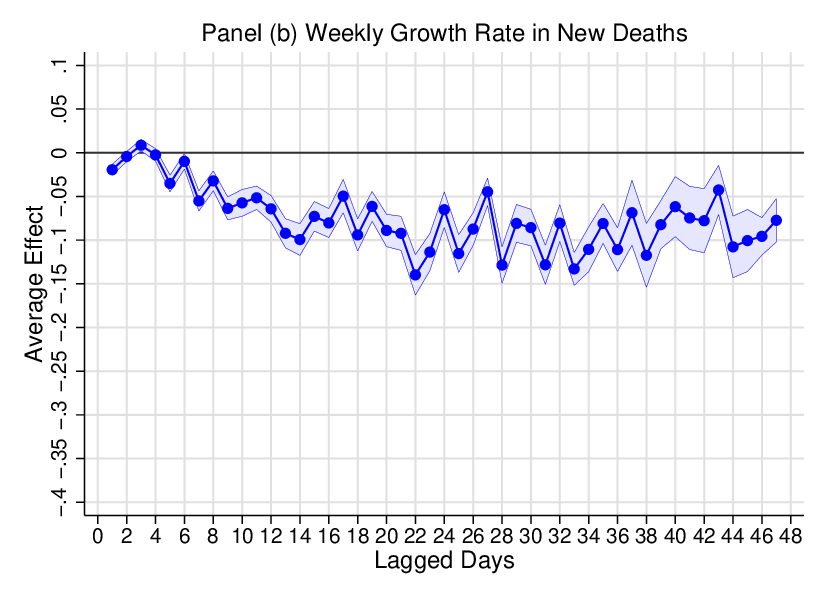}}
		\caption{\label{figure:att3} Average Treatment Effects of Social Distancing on Growth Rate of New Infections and COVID-Related Deaths Using an Extended Time Window.}
	\end{center}
\end{figure}

For the second robustness test, we extended our main analysis to include data up to May 7, 2020, allowing us to capture the dynamics of infection rates after the relaxation of social distancing measures. As presented in \Cref{figure:att3}, our findings indicate that there was indeed a rebound in cases after the lifting of stay-at-home orders. This provides further evidence supporting the effectiveness of these policies in controlling the spread of the virus. The extended analysis strengthens our study by addressing the potential longer-term impacts of stay-at-home orders and their subsequent relaxation.

\clearpage
\section{SEIR Model}\label{section:appendix_model}

Following the work of \cite{Kucharski2020}, we construct a modified SEIR model to investigate the transmission dynamics of COVID-19. Our model divides the total population into four distinct classes: susceptible, exposed, infectious, and removed (individuals who are isolated or recovered). 

We denote $S_{it}$ as the number of susceptible individuals in state $i$ at time $t$, $E_{it}^{1}$ and $E_{it}^{2}$ as the number of individuals in state $i$ during the first and second incubation periods, respectively, $I_{it}^{1}$ and $I_{it}^{2}$ as the number of individuals in state $i$ during the first and second infectious periods, respectively, $R_{it}$ as the number of recovered individuals in state $i$ at time $t$, $Q_{it}$ as the number of symptomatic individuals in state $i$ yet to be reported at time $t$, $C_{it}$ as the cumulative number of confirmed cases in state $i$ at time $t$, and $D_{it}$ as the cumulative number of confirmed deaths in state $i$ at time $t$. We assume that all individuals become symptomatic and infectious simultaneously and that the population in state $i$, $N_{i}$, remains constant. The law of motions for state $i$ is derived as follows:
\begin{align*} 
	S_{it+1}&=S_{it} - \beta_{t} S_{it} \left[I_{it}^{1}+I_{it}^{2}\right] / N_{i}\\
	E_{it+1}^{1}&=E_{it}^{1}+(1-f) \beta_{t} S_{it} \left[I_{it}^{1}+I_{it}^{2}\right] / N_{i}-2 \sigma E_{it}^{1}\\	 		 	
	E_{it+1}^{2}&=E_{it}^{2}+ 2 \sigma E_{it}^{1} - 2 \sigma E_{it}^{2}\\	 	
	I_{it+1}^{1}&=I_{it}^{1}+ 2 \sigma E_{it}^{2} - 2 \gamma I_{it}^{1}\\
	I_{it+1}^{2}&=I_{it}^{2}+ 2 \gamma I_{it}^{1} - 2 \gamma I_{it}^{2}\\
	Q_{it+1}&=Q_{it}+2 \sigma E_{it}^{2}\mathrm{e}^{-\gamma \kappa}-\kappa Q_{it}\\
	C_{it+1}&=C_{it}+ (1-\kappa) \eta  Q_{it}\\
	D_{it+1}&=D_{it}+\kappa \eta Q_{it}
\end{align*}
where $\beta_{t}$ represents the transmission rate at time $t$, $\sigma$ denotes the probability of becoming symptomatic, $\gamma$ is the probability of isolation, $\kappa$ is the probability of death conditional on being infectious, $\eta$ is the probability of recovery, and $f$ is the fraction of cases that travel to other states.

Furthermore, we model the law of motions in state $j$ as follows:
\begin{align*} 
	E_{jt+1}^{1}&=E_{jt}^{1}+f \beta_{t} S_{it} \left[I_{it}^{1}+I_{it}^{2}\right] / N_{i}-2 \sigma E_{jt}^{1}\\	 		 	
	E_{jt+1}^{2}&=E_{jt}^{2}+ 2 \sigma E_{jt}^{1} - 2 \sigma E_{jt}^{2}\\
	I_{jt+1}^{1}&=I_{jt}^{1}+ 2 \sigma E_{jt}^{2} - 2 \gamma I_{jt}^{1}\\
	I_{jt+1}^{2}&=I_{jt}^{2}+ 2 \gamma I_{jt}^{1} - 2 \gamma I_{jt}^{2}\\
	Q_{jt+1}&=Q_{t}+2 \sigma E_{jt}^{2}\mathrm{e}^{-\gamma \kappa}-\kappa Q_{jt}\\
	C_{jt+1}&=C_{jt}+ (1-\kappa)\eta Q_{jt}\\
	D_{jt+1}&=D_{jt}+\kappa \eta Q_{jt}
\end{align*}
where $E_{jt}^{1}$ and $E_{jt}^{2}$ represent the number of individuals who traveled from state $i$ to state $j$ during the first and second incubation periods, respectively, $I_{jt}^{1}$ and $I_{jt}^{2}$ denote the number of individuals who traveled from state $i$ to state $j$ during the first and second infectious periods, respectively, $Q_{jt}$ is the number of symptomatic cases among travelers from state $i$ to state $j$ yet to be reported at time $t$, $C_{jt}$ is the cumulative number of confirmed cases in state $j$ at time $t$, and $D_{jt}$ is the cumulative number of confirmed deaths in state $j$ at time $t$.

Our model operates under the assumption that the COVID-19 outbreak in the United States originated from a single infectious case in New York, after which the entire population became susceptible \citep{NBERw27102,NBERw27483,Ferretti2020}. We model the transmission rate using a geometric random walk process and employ the branching process to estimate the unknown parameters of interest. Our model incorporates delays in symptom appearance and case reporting by including transitions between reporting states and disease states \citep{Manski2021,Viner2020}. Furthermore, we explicitly account for the uncertainty in case and death observation by modeling the observed dynamics of new cases, symptom appearance of new cases, and reported confirmation of new cases \citep{Springborn2016}.

\clearpage
\section{Supplementary Details}\label{section:appendix_details}

We show the detailed statewide stay-at-home orders timeline in \Cref{table:date}.


\begin{longtable}[c]{lccc}
		\caption{\label{table:date}Statewide Stay-at-Home Orders Timeline. The data, collected from state government websites, is updated as of April 20, 2020. The Governor of Pennsylvania initially issued stay-at-home orders for some counties on March 23, and then extended these orders statewide on April 1.}\\
		\toprule
		\multicolumn{3}{c}{Statewide Stay-at-Home Orders Timeline}                                                \\* \midrule
		\endfirsthead
		\endhead
		\bottomrule
		\endfoot
		\endlastfoot
		State                & \multicolumn{1}{c}{Date Announced} & \multicolumn{1}{c}{Effective   Date} \\ \hline
		Alabama              & April   3                          & April 4                              \\
		Alaska               & March   27                         & March 28                             \\
		Arizona              & March   30                         & March 31                             \\
		Arkansas             & -                                  & -                                    \\
		California           & March   19                         & March 19                             \\
		Colorado             & March   26                         & March 26                             \\
		Connecticut          & March   20                         & March 23                             \\
		Delaware             & March   22                         & March 24                             \\
		District of Columbia & March   30                         & April 1                              \\
		Florida              & April   1                          & April 3                              \\
		Georgia              & April   2                          & April 3                              \\
		Hawaii               & March   23                         & March 25                             \\
		Idaho                & March   25                         & March 25                             \\
		Illinois             & March   20                         & March 21                             \\
		Indiana              & March   23                         & March 24                             \\
		Iowa                 & -                                  & -                                    \\
		Kansas               & March   28                         & March 30                             \\
		Kentucky             & March   22                         & March 26                             \\
		Louisiana            & March   22                         & March 23                             \\
		Maine                & March   31                         & April 2                              \\
		Maryland             & March   30                         & March 30                             \\
		Massachusetts        & March   23                         & March 24                             \\
		Michigan             & March   23                         & March 24                             \\
		Minnesota            & March   25                         & March 27                             \\
		Mississippi          & March   31                         & April 3                              \\
		Missouri             & April   3                          & April 6                              \\
		Montana              & March   26                         & March 28                             \\
		Nebraska             & -                                  & -                                    \\
		Nevada               & April   1                          & April   1                            \\
		New Hampshire        & March   26                         & March 27                             \\
		New Jersey           & March   20                         & March 21                             \\
		New Mexico           & March   23                         & March   24                           \\
		New York             & March   20                         & March 22                             \\
		North Carolina       & March   27                         & March 30                             \\
		North Dakota         & -                                  & -                                    \\
		Ohio                 & March   22                         & March 23                             \\
		Oklahoma             & -                                  & -                                    \\
		Oregon               & March   23                         & March 23                             \\
		Pennsylvania         & March   23                         & April   1                            \\
		Rhode Island         & March   28                         & March 28                             \\
		South Carolina       & -                                  & -                                    \\
		South Dakota         & -                                  & -                                    \\
		Tennessee            & March   30                         & March 31                             \\
		Texas                & March   31                         & April   2                            \\
		Utah                 & -                                  & -                                    \\
		Vermont              & March   24                         & March 24                             \\
		Virginia             & March   30                         & March 30                             \\
		Washington           & March   23                         & March 23                             \\
		West Virginia        & March   23                         & March 24                             \\
		Wisconsin            & March   24                         & March 25                             \\
		Wyoming              & -                                  & -                                    \\
		Guam              & -                                  & -                                    \\
		Northern  Mariana  Islands              & -                                  & -                                    \\
		Puerto Rico              & -                                  & -                                    \\
		Virgin Islands              & -                                  & -                                    \\* \bottomrule
	\end{longtable}

\end{document}